\documentclass[11pt]{article}
\usepackage{amssymb,amsmath,amsfonts}
\usepackage{graphicx}
\usepackage{graphics}
\usepackage{eepic,epsfig}

\@addtoreset{equation}{section}

\makeatother

\begin{document}

\title{Fermionic Casimir densities in a conical space with\\
a circular boundary and magnetic flux}
\author{E. R. Bezerra de Mello$^{1}$\thanks{%
E-mail: emello@fisica.ufpb.br},\thinspace\ F. Moraes$^{1}$\thanks{%
E-mail: moraes@fisica.ufpb.br}, \thinspace\ A. A. Saharian$^{1,2}$\thanks{%
E-mail: saharian@ysu.am} \\
\\
\textit{$^{1}$Departamento de F\'{\i}sica, Universidade Federal da Para\'{\i}%
ba}\\
\textit{58.059-970, Caixa Postal 5.008, Jo\~{a}o Pessoa, PB, Brazil}\vspace{%
0.3cm}\\
\textit{$^2$Department of Physics, Yerevan State University,}\\
\textit{1 Alex Manoogian Street, 0025 Yerevan, Armenia}}
\maketitle

\begin{abstract}
The vacuum expectation value (VEV) of the energy-momentum tensor for a
massive fermionic field is investigated in a (2+1)-dimensional conical
spacetime in the presence of a circular boundary and an infinitely thin
magnetic flux located at the cone apex. The MIT bag boundary condition is
assumed on the circle. At the cone apex we consider a special case of
boundary conditions for irregular modes, when the MIT bag boundary condition
is imposed at a finite radius, which is then taken to zero. The presence of
the magnetic flux leads to the Aharonov-Bohm-like effect on the VEV of the
energy-momentum tensor. For both exterior and interior regions, the VEV is
decomposed into boundary-free and boundary-induced parts. Both these parts
are even periodic functions of the magnetic flux with the period equal to
the flux quantum. The boundary-free part in the radial stress is equal to
the energy density. Near the circle, the boundary-induced part in the VEV
dominates and for a massless field the vacuum energy density is negative
inside the circle and positive in the exterior region. Various special cases
are considered.
\end{abstract}

\bigskip


\bigskip

\section{Introduction}

Topological defects created after symmetry-breaking phase transitions play
an important role in many fields of physics (for a review see \cite{Vile94}%
). They appear in different condensed matter systems including superfluids,
superconductors, and liquid crystals. Moreover, the topological defects
provide an important link between particle physics and cosmology. In
particular, the cosmic strings are candidates to produce a number of
interesting physical effects, such as the generation of gravitational waves,
gamma ray bursts, and high-energy cosmic rays. In the simplest theoretical
model, the geometry of a cosmic string outside its core is described by the
planar angle deficit. Though this geometry is flat, the corresponding
nontrivial topology leads to a number of interesting physical effects. In
particular, the properties of the quantum vacuum are changed due to the
modification of the zero-point fluctuations spectrum of quantum fields.
Explicit calculations for the vacuum polarization by the cosmic string have
been developed for different fields and in various spatial dimensions (see,
for instance, \cite{Hell86}-\cite{Beze06}). The combined effects of the
cosmic string topology and a coaxial cylindrical boundary on the
polarization of the vacuum were studied in Refs. \cite{Brev95}-\cite%
{Beze08Ferm} for scalar, electromagnetic and fermionic fields.

In Refs. \cite{Beze10b,Bell11} we have investigated the vacuum expectation
value (VEV) of the fermionic current and the fermionic condensate induced by
the vortex configuration of a gauge field in a (2+1)-dimensional conical
space with a circular boundary. Continuing in this line of investigation, in
the present paper we study the VEV of the energy-momentum tensor for a
massive fermionic field with the MIT bag boundary condition. The imposition
of the boundary condition induces sifts in the expectation values of
physical characteristics of the vacuum state. This is the well known Casimir
effect \cite{Casimir}. The expectation value of the energy-momentum tensor
is among the most important characteristics of the vacuum. In addition to
describing the physical structure of a quantum field at a given point, it
acts as the source of gravity in the quasiclassical Einstein equations and
plays an important role in modelling self-consistent dynamics involving the
gravitational field. In considering the expectation value of the
energy-momentum tensor we shall assume the presence of a magnetic flux. The
interaction of a magnetic flux tube with a fermionic field gives rise to a
number of interesting phenomena, such as the Aharonov-Bohm effect, parity
anomalies, formation of a condensate, and generation of exotic quantum
numbers. For background Minkowski spacetime, the combined effects of the
magnetic flux and boundaries on the vacuum energy have been studied in Refs.
\cite{Lese98,Bene00}.

Field theories in 2+1 dimensions provide simple models in particle physics.
Related theories also arise in the long-wavelength description of certain
planar condensed matter systems, including models of high-temperature
superconductivity. They exhibit a number of interesting effects, such as
parity violation, flavor symmetry breaking, and fractionalization of quantum
numbers (see Refs. \cite{Dese82}-\cite{Jack09}). An important aspect is the
possibility of giving a topological mass to the gauge bosons without
breaking gauge invariance. An interesting application of Dirac theory in 2+1
dimensions recently appeared in nanophysics. The long-wavelength description
of the electronic states in a graphene sheet can be formulated in terms of
the Dirac-like theory of massless spinors in (2+1)-dimensional spacetime
with the Fermi velocity playing the role of the speed of light (for a review
see Ref. \cite{Cast09}). One-loop quantum effects induced by the nontrivial
topology of graphene made cylindrical and toroidal nanotubes have been
recently considered in Ref. \cite{Bell09}. The vacuum polarization in
graphene with a topological defect is investigated in Ref. \cite{Site08}
within the framework of a long-wavelength continuum model.

The outline of the paper is as follows. In the next section we consider the
geometry of a boundary-free conical space with an infinitesimally thin
magnetic flux placed at the apex of the cone. At the cone apex, a special
case of boundary conditions is considered, when the MIT bag boundary
condition is imposed at a finite radius, which is then taken to zero. The
renormalized VEV of the energy-momentum tensor is evaluated. Integral
representations are provided for the energy density and vacuum stresses. In
Sect. \ref{sec:EMTinside}, we consider the VEV of the energy-momentum tensor
in the region inside a circular boundary with the MIT bag boundary
condition. The VEV is decomposed into boundary-free and boundary induced
parts. A rapidly convergent integral representation for the latter is
obtained. A similar investigation for the region outside a circular boundary
is presented in Sect. \ref{sec:EMToutside}. The case with half-integer
values of the ratio of the magnetic flux to the flux quantum requires a
special consideration. The corresponding analysis is presented in Sect. \ref%
{sec:EMTspecial}. Finally, Sect. \ref{sec:Conc} contains a summary of the
work.

\section{Energy-momentum tensor in a boundary-free conical space}

\label{sec:BoundFree}

In the presence of the external electromagnetic field with the vector
potential $A_{\mu }$, the dynamics of a massive spinor field $\psi $ is
governed by the Dirac equation
\begin{equation}
i\gamma ^{\mu }(\nabla _{\mu }+ieA_{\mu })\psi -m\psi =0\ ,  \label{Direq}
\end{equation}%
with $\gamma ^{\mu }=e_{(a)}^{\mu }\gamma ^{(a)}$ being the Dirac matrices.
Here $\gamma ^{(a)}$ are the flat spacetime gamma matrices and $e_{(a)}^{\mu
}$ is the basis tetrad. The covariant derivative operator is given by the
relation
\begin{equation}
\nabla _{\mu }=\partial _{\mu }+\frac{1}{4}\gamma ^{(a)}\gamma
^{(b)}e_{(a)}^{\nu }e_{(b)\nu ;\mu }\ ,  \label{Gammamu}
\end{equation}%
where "$;$" means the standard covariant derivative for vector fields. As a
background geometry, we consider a $(2+1)$-dimensional conical spacetime
with the line element
\begin{equation}
ds^{2}=g_{\mu \nu }dx^{\mu }dx^{\nu }=dt^{2}-dr^{2}-r^{2}d\phi ^{2},
\label{ds21}
\end{equation}%
with $r\geqslant 0$ and $0\leqslant \phi \leqslant \phi _{0}$. In $(2+1)$%
-dimensional spacetime there are two inequivalent irreducible
representations of the Clifford algebra. In what follows we choose the flat
space Dirac matrices in the form $\gamma ^{(0)}=\sigma _{3}$, $\gamma
^{(1)}=i\sigma _{1}$, $\gamma ^{(2)}=i\sigma _{2}$, where $\sigma _{l}$ are
Pauli matrices. In the second representation the gamma matrices can be taken
as $\gamma ^{(0)}=-\sigma _{3}$, $\gamma ^{(1)}=-i\sigma _{1}$, $\gamma
^{(2)}=-i\sigma _{2}$. The corresponding results for the second
representation are obtained by changing the sign of the mass, $m\rightarrow
-m$.

We assume the presence of a circular boundary with radius $a$ on which the
field obeys the MIT bag boundary condition
\begin{equation}
\left( 1+in_{\mu }\gamma ^{\mu }\right) \psi \big|_{r=a}=0\ ,  \label{BCMIT}
\end{equation}%
where $n_{\mu }$ is the outward directed normal (with respect to the region
under consideration) to the boundary. We have $n_{\mu }=\delta _{\mu }^{1}$
and $n_{\mu }=-\delta _{\mu }^{1}$ for the interior and exterior regions
respectively. As it will be shown below, the VEV is decomposed into the
boundary-free and boundary induced parts. In this section, we shall be
concerned with the VEV of the energy-momentum tensor operator for a spinor
field in the boundary-free conical space. We assume the magnetic field
configuration corresponding to an infinitely thin magnetic flux located at
the apex of the cone. In the cylindrical coordinates of Eq. (\ref{ds21}),
the corresponding vector potential has the components $A_{\mu }=(0,0,A)$ for
$r>0$. The $z$-component is related to the magnetic flux $\Phi $ by the
formula $A=-\Phi /\phi _{0}$.

For the evaluation of the VEV of the energy-momentum tensor we use the
mode-sum formula
\begin{equation}
\left\langle 0\left\vert T_{\mu \nu }\right\vert 0\right\rangle =\frac{i}{2}%
\sum_{\sigma }\left[ \bar{\psi}_{\sigma }^{(-)}(x)\gamma _{(\mu }\nabla
_{\nu )}\psi _{\sigma }^{(-)}(x)-(\nabla _{(\mu }\bar{\psi}_{\sigma
}^{(-)}(x))\gamma _{\nu )}\psi _{\sigma }^{(-)}(x)\right] \ .
\label{modesum}
\end{equation}%
where $\{\psi _{\sigma }^{(+)},\psi _{\sigma }^{(-)}\}$ is a complete set of
positive- and negative-energy solutions to the Dirac equation, $\bar{\psi}%
=\psi ^{\dagger }\gamma ^{0}$ is the Dirac adjoint and the dagger denotes
Hermitian conjugation. Here $\sigma $ stands for a set of quantum numbers
specifying the solutions (see below). The theory of von Neumann deficiency
indices leads to a one-parameter family of allowed boundary conditions in
the background of an Aharonov-Bohm gauge field \cite{Sous89}. Here we
consider a special case of boundary conditions at the cone apex, when the
MIT bag boundary condition is imposed at a finite radius, which is then
taken to zero. The VEVs for other boundary conditions are evaluated in a
similar way. The contribution of the regular modes is the same for all
boundary conditions and the results differ by the parts related to the
irregular modes.

The mode functions in the boundary-free conical space are specified by the
set $\sigma =(\gamma ,j)$ with $0\leqslant \gamma <\infty $ and $j=\pm
1/2,\pm 3/2,\ldots $. The corresponding normalized negative-energy
eigenspinors are given by the expression \cite{Beze10b}%
\begin{equation}
\psi _{(0)\gamma j}^{(-)}=\left( \gamma \frac{E+m}{2\phi _{0}E}\right)
^{1/2}e^{-iqj\phi +iEt}\left(
\begin{array}{c}
\frac{\gamma \epsilon _{j}e^{-iq\phi /2}}{E+m}J_{\beta _{j}+\epsilon
_{j}}(\gamma r) \\
J_{\beta _{j}}(\gamma r)e^{iq\phi /2}%
\end{array}%
\right) ,  \label{psi0}
\end{equation}%
where $E=\sqrt{\gamma ^{2}+m^{2}}$ and $J_{\nu }(x)$ is the Bessel function.
In Eq. (\ref{psi0}) we have defined%
\begin{equation}
\beta _{j}=q|j+\alpha |-\epsilon _{j}/2,\;q=2\pi /\phi _{0},  \label{jbetj}
\end{equation}%
with%
\begin{equation}
\alpha =eA/q=-e\Phi /2\pi ,  \label{alfatilde}
\end{equation}%
and
\begin{equation}
\epsilon _{j}=\left\{
\begin{array}{cc}
1, & \;j>-\alpha \\
-1, & \;j<-\alpha%
\end{array}%
\right. .  \label{epsj}
\end{equation}%
The expression for the positive-energy eigenspinor is found from Eq. (\ref%
{psi0}) by using the relation $\psi _{\gamma j}^{(+)}=\sigma _{1}\psi
_{\gamma j}^{(-)\ast }$, where the asterisk means complex conjugate. Here we
assume that the parameter $\alpha $ is not a half-integer. The special case
of half-integer $\alpha $ will be considered separately in Sect. \ref%
{sec:EMTspecial}.

Substituting the eigenspinors (\ref{psi0}) into the mode-sum (\ref{modesum}%
), for the VEV of the energy-momentum tensor in the boundary-free geometry, $%
\left\langle 0\left\vert T_{\mu }^{\nu }\right\vert 0\right\rangle =\langle
T_{\mu }^{\nu }\rangle _{0}$, one finds%
\begin{eqnarray}
\langle T_{0}^{0}\rangle _{0} &=&-\frac{q}{4\pi }\sum_{j}\int_{0}^{\infty
}d\gamma \,\gamma \lbrack \left( E-m\right) J_{\beta _{j}+\epsilon
_{j}}^{2}(\gamma r)+\left( E+m\right) J_{\beta _{j}}^{2}(\gamma r)],  \notag
\\
\langle T_{1}^{1}\rangle _{0} &=&\frac{q}{4\pi }\sum_{j}\epsilon
_{j}\int_{0}^{\infty }d\gamma \,\frac{\gamma ^{3}}{E}[J_{\beta _{j}}(\gamma
r)J_{\beta _{j}+\epsilon _{j}}^{\prime }(\gamma r)-J_{\beta _{j}}^{\prime
}(\gamma r)J_{\beta _{j}+\epsilon _{j}}(\gamma r)],  \label{Tii0} \\
\langle T_{2}^{2}\rangle _{0} &=&\frac{q}{4\pi r}\sum_{j}\epsilon
_{j}\int_{0}^{\infty }d\gamma \,\frac{\gamma ^{2}}{E}\left( 2\epsilon
_{j}\beta _{j}+1\right) J_{\beta _{j}}(\gamma r)J_{\beta _{j}+\epsilon
_{j}}(\gamma r),  \notag
\end{eqnarray}%
where $\sum_{j}$ means the summation over $j=\pm 1/2,\pm 3/2,\ldots $ and
the prime means the derivative with respect to the argument of the function.
By using the relation
\begin{equation}
J_{\beta _{j}}^{\prime }(z)=-\epsilon _{j}J_{\beta _{j}+\epsilon
_{j}}(z)+\epsilon _{j}\frac{\beta _{j}}{z}J_{\beta _{j}}(z),  \label{BessRel}
\end{equation}%
the VEVs of the energy density and radial stress may be written in the form%
\begin{equation}
\langle T_{0}^{0}\rangle _{0}=-A_{0}(r)+m\langle \bar{\psi}\psi \rangle
_{0},\;\langle T_{1}^{1}\rangle _{0}=A_{0}(r)-\langle T_{2}^{2}\rangle _{0},
\label{T0011}
\end{equation}%
with $\langle \bar{\psi}\psi \rangle _{0}$ being the fermionic condensate
(see Ref. \cite{Bell11}) and%
\begin{equation}
A_{0}(r)=\frac{q}{4\pi }\sum_{j}\int_{0}^{\infty }d\gamma \,\frac{\gamma ^{3}%
}{E}\left[ J_{\beta _{j}}^{2}(\gamma r)+J_{\beta _{j}+\epsilon
_{j}}^{2}(\gamma r)\right] .  \label{A0}
\end{equation}%
From these expressions the trace relation $\langle T_{k}^{k}\rangle
_{0}=m\langle \bar{\psi}\psi \rangle _{0}$ is explicitly seen. Another
relation between the separate components is a consequence of the covariant
conservation equation for the energy-momentum tensor. For the geometry at
hand the latter is reduced to a single equation: $\partial _{r}\left(
r\langle T_{1}^{1}\rangle _{0}\right) =\langle T_{2}^{2}\rangle _{0}$.

If we present the parameter $\alpha $ related to the magnetic flux as
\begin{equation}
\alpha =\alpha _{0}+n_{0},\;|\alpha _{0}|<1/2,  \label{alf}
\end{equation}%
with $n_{0}$ being an integer number, it can be seen that the VEVs do not
depend on $n_{0}$. Hence, we conclude that the VEV of the energy-momentum
tensor depends on $\alpha _{0}$ alone. The VEV is an even function of this
parameter (note that the same is the case for the fermionic condensate,
whereas the VEV of the fermionic current is an odd function of $\alpha _{0}$%
).

The expressions (\ref{Tii0}) are divergent and need to be regularized. We
introduce a cutoff function $e^{-s\gamma ^{2}}$ with the cutoff parameter $%
s>0$. At the end of calculations the limit $s\rightarrow 0$ is taken. First
we consider the azimuthal stress. From Eq. (\ref{Tii0}), the corresponding
regularized VEV can be written in the form%
\begin{equation}
\langle T_{2}^{2}\rangle _{0,\text{reg}}=\frac{q}{8\pi r^{2}}\sum_{j}\left(
2\beta _{j}+\epsilon _{j}\right) \left( 2\beta _{j}-\epsilon _{j}r\partial
_{r}\right) \int_{0}^{\infty }d\gamma \,\frac{\gamma e^{-s\gamma ^{2}}}{%
\sqrt{\gamma ^{2}+m^{2}}}J_{\beta _{j}}^{2}(\gamma r).  \label{T22reg}
\end{equation}%
By using the relation
\begin{equation}
\frac{1}{\sqrt{\gamma ^{2}+m^{2}}}=\frac{2}{\sqrt{\pi }}\int_{0}^{\infty
}dte^{-(\gamma ^{2}+m^{2})t^{2}},  \label{repres}
\end{equation}%
and changing the order of integrations, the $\gamma $-integral is performed
explicitly (see Ref. \cite{Prud86}) with the result%
\begin{eqnarray}
\langle T_{2}^{2}\rangle _{0,\text{reg}} &=&\frac{q}{8\pi r^{2}}\frac{%
e^{m^{2}s}}{\sqrt{2\pi }}\sum_{j}\left( 2\beta _{j}+\epsilon _{j}\right)
\left( 2\beta _{j}-\epsilon _{j}r\partial _{r}\right)  \notag \\
&&\times \int_{0}^{1/(2s)}dy\frac{y^{-1/2}I_{\beta _{j}}\left( r^{2}y\right)
}{\sqrt{1-2ys}}\,e^{-m^{2}/(2y)-r^{2}y},  \label{T22reg1}
\end{eqnarray}%
where $I_{\beta _{j}}\left( x\right) $ is the modified Bessel function. By
using the properties of the modified Bessel function, Eq. (\ref{T22reg1})
can also be written in the form%
\begin{equation}
\langle T_{2}^{2}\rangle _{0,\text{reg}}=\frac{qe^{m^{2}s}}{2(2\pi )^{3/2}}%
\partial _{r}r\int_{0}^{1/(2s)}dy\frac{y^{1/2}\,e^{-m^{2}/(2y)-r^{2}y}}{%
\sqrt{1-2ys}}\sum_{j}\left[ I_{\beta _{j}}(r^{2}y)+I_{\beta _{j}+\epsilon
_{j}}(r^{2}y)\right] .  \label{T22reg2}
\end{equation}

In order to find the representation for regularized VEVs of the energy
density and radial stress we need to consider the expression for $A_{0}(r)$
in Eq. (\ref{A0}) regularized with the cutoff function $e^{-s\gamma ^{2}}$.
The regularized expression can be presented in the form%
\begin{equation}
A_{0,\text{reg}}(r)=-\frac{q}{4\pi }\sum_{j}\partial _{s}\int_{0}^{\infty
}d\gamma \,\frac{\gamma e^{-s\gamma ^{2}}}{\sqrt{\gamma ^{2}+m^{2}}}\left[
J_{\beta _{j}}^{2}(\gamma r)+J_{\beta _{j}+\epsilon _{j}}^{2}(\gamma r)%
\right] .  \label{A0reg}
\end{equation}%
The parts with separate terms in the square brackets are evaluated in a way
similar to that we used for Eq. (\ref{T22reg}). As a result we find%
\begin{equation}
A_{0,\text{reg}}(r)=\frac{qe^{m^{2}s}}{(2\pi )^{3/2}}\sum_{j}\frac{\partial
}{\partial r^{2}}r^{2}\int_{0}^{1/(2s)}dy\frac{y^{1/2}e^{-m^{2}/(2y)-r^{2}y}%
}{\sqrt{1-2ys}}\left[ I_{\beta _{j}}\left( r^{2}y\right) +I_{\beta
_{j}+\epsilon _{j}}\left( r^{2}y\right) \right] .  \label{A0reg1}
\end{equation}%
From here, in the combination with Eq. (\ref{T0011}), for the regularized
radial stress we obtain the expression%
\begin{equation}
\langle T_{1}^{1}\rangle _{0,\text{reg}}=\frac{qe^{m^{2}s}}{2(2\pi )^{3/2}}%
\int_{0}^{1/(2s)}dy\frac{y^{1/2}e^{-m^{2}/(2y)-r^{2}y}}{\sqrt{1-2ys}}\sum_{j}%
\left[ I_{\beta _{j}}\left( r^{2}y\right) +I_{\beta _{j}+\epsilon
_{j}}\left( r^{2}y\right) \right] .  \label{T11reg}
\end{equation}%
For the energy density and the azimuthal stress we have:%
\begin{eqnarray}
\langle T_{0}^{0}\rangle _{0,\text{reg}} &=&-\left( 2+r\partial _{r}\right)
\langle T_{1}^{1}\rangle _{0,\text{reg}}+m\langle \bar{\psi}\psi \rangle _{0,%
\text{reg}},  \notag \\
\langle T_{2}^{2}\rangle _{0,\text{reg}} &=&\left( 1+r\partial _{r}\right)
\langle T_{1}^{1}\rangle _{0,\text{reg}}.  \label{T0022}
\end{eqnarray}%
Note that by the second relation we explicitly checked the covariant
continuity equation for the regularized VEVs.

As the fermionic condensate has been considered in Ref. \cite{Bell11}, in
accordance with Eq. (\ref{T0022}) we need to consider the radial stress
only. The corresponding regularized VEV is expressed in terms of the series%
\begin{equation}
\mathcal{I}(q,\alpha ,z)=\sum_{j}I_{\beta _{j}}(z).  \label{Iser}
\end{equation}%
If we present the parameter $\alpha $ in the form (\ref{alf}), it is easily
seen the independence of the series on $n_{0}$: $\mathcal{I}(q,\alpha ,z)=%
\mathcal{I}(q,\alpha _{0},z)$. For the second series appearing in the
expressions for the regularized VEVs we have
\begin{equation}
\sum_{j}I_{\beta _{j}+\epsilon _{j}}(z)=\mathcal{I}(q,-\alpha _{0},z).
\label{Iser1}
\end{equation}

With the notation (\ref{Iser}), the regularized VEV of the radial stress is
written in the form%
\begin{equation}
\langle T_{1}^{1}\rangle _{0,\text{reg}}=\frac{qe^{m^{2}s}}{2(2\pi )^{3/2}}%
\int_{0}^{1/(2s)}dy\frac{y^{1/2}e^{-m^{2}/(2y)-r^{2}y}}{\sqrt{1-2ys}}%
\sum_{j=\pm 1}\mathcal{I}(q,j\alpha _{0},r^{2}y).  \label{T22reg3}
\end{equation}%
For $2p<q<2p+2$, with $p$ being an integer, we use the representation \cite%
{Beze10b}%
\begin{equation}
\mathcal{I}(q,\alpha _{0},z)=\frac{e^{z}}{q}+\mathcal{J}(q,\alpha _{0},z),
\label{Iser2}
\end{equation}%
with the notation
\begin{eqnarray}
\mathcal{J}(q,\alpha _{0},z) &=&-\frac{1}{\pi }\int_{0}^{\infty }dy\frac{%
e^{-z\cosh y}f(q,\alpha _{0},y)}{\cosh (qy)-\cos (q\pi )}  \notag \\
&&+\frac{2}{q}\sum_{l=1}^{p}(-1)^{l}\cos [2\pi l(\alpha _{0}-1/2q)]e^{z\cos
(2\pi l/q)}.  \label{Iser4}
\end{eqnarray}%
The function in the integrand is defined by the expression%
\begin{eqnarray}
f(q,\alpha _{0},y) &=&\cos \left[ q\pi \left( 1/2-\alpha _{0}\right) \right]
\cosh \left[ \left( q\alpha _{0}+q/2-1/2\right) y\right]  \notag \\
&&-\cos \left[ q\pi \left( 1/2+\alpha _{0}\right) \right] \cosh \left[
\left( q\alpha _{0}-q/2-1/2\right) y\right] .  \label{f}
\end{eqnarray}%
In the case $q=2p$, the term
\begin{equation}
-(-1)^{q/2}\frac{e^{-z}}{q}\sin (q\pi \alpha _{0}),  \label{q2p}
\end{equation}%
should be added to the right-hand side of Eq. (\ref{Iser4}). For $1\leqslant
q<2$, the last term on the right-hand side of Eq. (\ref{Iser4}) is absent.

Substituting Eq. (\ref{Iser2}) into the right-hand side of Eq. (\ref{T22reg3}%
), we can see that the only divergent contribution comes from the term $%
e^{z}/q$. This contribution does not depend on the opening angle of the cone
and on the magnetic flux. It coincides with the corresponding quantity in
the Minkowski spacetime, in the absence of the magnetic flux. Subtracting
the Minkowskian part and taking the limit $s\rightarrow 0$, after the
explicit integration over $y$, we get the expression for the renormalized
radial stress, $\langle T_{1}^{1}\rangle _{0,\text{ren}}$. The corresponding
expressions for the energy density and the azimuthal stress are found from
Eq. (\ref{T0022}), by using the expression for $\langle \bar{\psi}\psi
\rangle _{0,\text{ren}}$ from Ref. \cite{Bell11}. In this way, one finds the
following formula (no summation over $i$)%
\begin{eqnarray}
\langle T_{i}^{i}\rangle _{0,\text{ren}} &=&\frac{m^{3}}{\pi }\left[
\sum_{l=1}^{p}(-1)^{l}\cos (\pi l/q)\cos (2\pi l\alpha
_{0})F_{i}^{(s)}(2mrs_{l})\right.  \notag \\
&&\left. -\frac{q}{2\pi }\int_{0}^{\infty }dy\frac{\sum_{\delta =\pm
1}f(q,\delta \alpha _{0},2y)F_{i}^{(s)}(2mr\cosh (y))}{\cosh (2qy)-\cos
(q\pi )}\right] ,  \label{T11ren}
\end{eqnarray}%
where $p$ is an integer defined by $2p\leqslant q<2p+2$, and
\begin{equation}
s_{l}=\sin (\pi l/q).  \label{sl}
\end{equation}%
In Eq. (\ref{T11ren}), we have defined the functions%
\begin{eqnarray}
F_{0}^{(s)}(u) &=&F_{1}^{(s)}(u)=\frac{e^{-u}}{u^{3}}\left( u+1\right) ,
\notag \\
F_{2}^{(s)}(u) &=&-\frac{e^{-u}}{u^{3}}\left( u^{2}+2u+2\right) .
\label{Fsi}
\end{eqnarray}%
Note that for the function in the integrand one has:
\begin{equation}
\sum_{\delta =\pm 1}f(q,\delta \alpha _{0},2y)=-2\sinh (y)\sum_{\delta =\pm
1}\cos \left[ q\pi \left( 1/2+\delta \alpha _{0}\right) \right] \sinh
[q\left( 1-2\delta \alpha _{0}\right) y].  \label{fsum}
\end{equation}%
As it is seen, the radial stress is equal to the energy density: $\langle
T_{0}^{0}\rangle _{0,\text{ren}}=\langle T_{1}^{1}\rangle _{0,\text{ren}}$.

\begin{figure}[tbph]
\begin{center}
\epsfig{figure=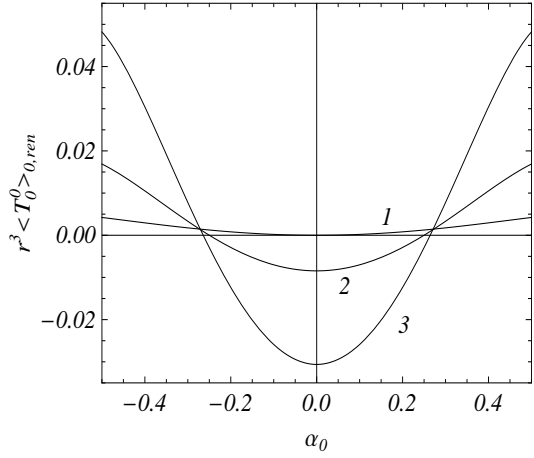,width=7.cm,height=6.cm}
\end{center}
\caption{Energy density for a massless fermionic field as a function of the
parameter $\protect\alpha _{0}$ for separate values of the parameter $q$
(numbers near the curves). The vacuum stresses are related to the energy
density by Eq. (\protect\ref{Fism0}).}
\label{fig1}
\end{figure}

For a massless field the corresponding energy density is directly obtained
from Eq. (\ref{T11ren}):%
\begin{eqnarray}
\langle T_{0}^{0}\rangle _{0,\text{ren}} &=&\frac{1}{8\pi r^{3}}\left[
\sum_{l=1}^{p}\frac{(-1)^{l}}{s_{l}^{3}}\cos (\pi l/q)\cos (2\pi l\alpha
_{0})\right.  \notag \\
&&\left. -\frac{q}{2\pi }\int_{0}^{\infty }dy\frac{\sum_{j=\pm 1}f(q,j\alpha
_{0},2y)}{\cosh (2qy)-\cos (q\pi )}\frac{1}{\cosh ^{3}y}\right] .
\label{Tiirenm0}
\end{eqnarray}%
For the radial and azimuthal stresses one has
\begin{equation}
\langle T_{1}^{1}\rangle _{0,\text{ren}}=-\frac{1}{2}\langle
T_{2}^{2}\rangle _{0,\text{ren}}=\langle T_{0}^{0}\rangle _{0,\text{ren}}.
\label{Fism0}
\end{equation}%
Of course, in the massless case the energy-momentum tensor is traceless. In
Fig. \ref{fig1} we plot the renormalized energy density for a massless field
as a function of the parameter $\alpha _{0}$ for separate values of the
parameter $q$ (numbers near the curves).

For a massive field, the expression in the right-hand side of Eq. (\ref%
{Tiirenm0}) gives the leading term in the corresponding asymptotic expansion
for small distances from the string, $mr\ll 1$. At distances larger than the
Compton wavelength of the spinor particle, $mr\gg 1$, the VEVs are
suppressed by the factor $e^{-2mr}$ for $1\leqslant q\leqslant 2$ and by the
factor $e^{-2mr\sin (\pi /q)}$ for $q>2$. In the latter case the dominant
contribution comes from the first term in the right-hand side of Eq. (\ref%
{T11ren}):%
\begin{eqnarray}
\langle T_{i}^{i}\rangle _{0,\text{ren}} &\approx &-\frac{m^{2}}{2\pi r}\cot
(\pi /q)\cos (2\pi \alpha _{0})  \notag \\
&&\times e^{-2mr\sin (\pi /q)}\left\{
\begin{array}{cc}
1/[2mr\sin (\pi /q)], & i=0,1 \\
-1, & i=2%
\end{array}%
\right. ,  \label{FClargem}
\end{eqnarray}%
for $mr\gg 1$.

For integer $q$ and for the parameter $\alpha $ given by the special value
\begin{equation}
\alpha =1/2q-1/2,  \label{alphaSpecial}
\end{equation}%
the expression (\ref{T11ren}) for the VEVs take the form (no summation)%
\begin{equation}
\langle T_{i}^{i}\rangle _{0,\text{ren}}=\frac{m^{3}}{2\pi }%
\sum_{l=1}^{q-1}\cos ^{2}(\pi l/q)F_{i}^{(s)}(2mrs_{l}),  \label{Tii0ren}
\end{equation}%
Note that, in this case, the renormalized VEV vanishes in a conical space
with $q=2$. For $q\geqslant 3$ the energy density is positive.

The energy density diverges on the string as $1/r^{3}$ and, as a result, the
integrated energy diverges as well. We can evaluate the total vacuum energy
in the region $r_{0}\leqslant r<\infty $ by using the energy density given
above: $E_{0,r\geqslant r_{0}}=\phi _{0}\int_{r_{0}}^{\infty }dr\,r\langle
T_{0}^{0}\rangle _{0,\text{ren}}$. Performing the radial integration, we find%
\begin{eqnarray}
E_{0,r\geqslant r_{0}} &=&\frac{1}{4qr_{0}}\left[ \sum_{l=1}^{p}\frac{%
(-1)^{l}}{s_{l}^{3}}\cos (\pi l/q)\cos (2\pi l\alpha
_{0})e^{-2mr_{0}s_{l}}\right.  \notag \\
&&\left. -\frac{q}{2\pi }\int_{0}^{\infty }dy\frac{\sum_{j=\pm 1}f(q,j\alpha
_{0},2y)}{\cosh (2qy)-\cos (q\pi )}\frac{e^{-2mr_{0}\cosh y}}{\cosh ^{3}y}%
\right] .  \label{TotEn}
\end{eqnarray}%
For a massless field we have a simple relation $E_{0,r\geqslant r_{0}}=\phi
_{0}r_{0}^{2}\langle T_{0}^{0}\rangle _{0,\text{ren}}|_{r=r_{0}}$, which
could also be directly obtained from Eq. (\ref{Tiirenm0}).

In the special case when the magnetic flux is absent we have $\alpha _{0}=0$
and the general formula simplifies to (no summation over $i$)%
\begin{eqnarray}
\langle T_{i}^{i}\rangle _{0,\text{ren}} &=&\frac{m^{3}}{\pi }\left[
\sum_{l=1}^{p}(-1)^{l}\cos (\pi l/q)F_{i}^{(s)}(2mrs_{l})\right.  \notag \\
&&\left. +\frac{2q}{\pi }\cos \left( \frac{q\pi }{2}\right) \int_{0}^{\infty
}dy\frac{F_{i}^{(s)}(2mr\cosh y)\sinh (qy)\sinh y}{\cosh (2qy)-\cos (q\pi )}%
\right] .  \label{Tiialf0}
\end{eqnarray}%
In this case, the VEV is only a consequence of the conical structure of the
space. For odd values of the parameter $q$ the second term in the square
brackets vanishes and for the VEV we have the simple formula (no summation
over $i$)%
\begin{equation}
\langle T_{i}^{i}\rangle _{0,\text{ren}}=\frac{m^{3}}{\pi }%
\sum_{l=1}^{p}(-1)^{l}\cos (\pi l/q)F_{i}^{(s)}(2mrs_{l}).  \label{Tii0Oddq}
\end{equation}%
Another special case corresponds to the magnetic flux in background of
Minkowski spacetime. In this case, taking $q=1$, from the general formulas
we find%
\begin{equation}
\langle T_{i}^{i}\rangle _{0,\text{ren}}=\frac{m^{3}}{\pi ^{2}}\sin \left(
\pi \alpha _{0}\right) \int_{0}^{\infty }dy\tanh (y)\sinh \left( 2\alpha
_{0}y\right) F_{i}^{(s)}(2mr\cosh (y)),  \label{Tii0q1}
\end{equation}%
and the corresponding energy density is positive for $\alpha _{0}\neq 0$.

An alternative expression for the VEV of the energy-momentum tensor is
obtained by using the formula \cite{Beze10b}
\begin{eqnarray}
&&\mathcal{I}(q,\alpha _{0},x)=\frac{2}{q}\int_{0}^{\infty
}dz\,I_{z}(x)+A(q,\alpha _{0},x)\,  \notag \\
&&\qquad -\frac{4}{\pi q}\int_{0}^{\infty }dz\,{\mathrm{Re}}\left[ \frac{%
\sinh (z\pi )K_{iz}(x)}{e^{2\pi (z+i|q\alpha _{0}-1/2|)/q}+1}\right] ,
\label{Rep2}
\end{eqnarray}%
with $K_{\nu }(x)$ being the modified Bessel function. In Eq. (\ref{Rep2}), $%
A(q,\alpha _{0},x)=0$ for $|\alpha _{0}-1/2q|\leqslant 1/2$, and
\begin{equation}
A(q,\alpha _{0},x)=\frac{2}{\pi }\sin [\pi (|q\alpha
_{0}-1/2|-q/2)]K_{|q\alpha _{0}-1/2|-q/2}(x),  \label{Aq}
\end{equation}%
for $1/2<|\alpha _{0}-1/2q|<1$. Substituting the representation (\ref{Rep2})
into the expressions for the regularized VEVs, we see that the part with the
first term on the right-hand side of Eq. (\ref{Rep2}) does not depend on the
opening angle of the cone and on the magnetic flux. This term coincides with
the corresponding result in Minkowski bulk when the magnetic flux is absent.
Hence, it should be subtracted in the renormalization procedure. In the
remaining part the limit $s\rightarrow 0$ can be taken directly.

In a same way, we can consider a more general problem where the spinor field
obeys quasiperiodic boundary condition along the azimuthal direction%
\begin{equation}
\psi (t,r,\phi +\phi _{0})=e^{2\pi i\chi }\psi (t,r,\phi ),  \label{PerBC}
\end{equation}%
with a constant parameter $\chi $, $|\chi |\leqslant 1/2$. For this problem,
the exponential factor in the expression for the mode functions (\ref{psi0})
has the form $e^{-iq(n+\chi )\phi +iEt}$. The corresponding expression for
the mode functions is obtained from that given above with the parameter $%
\alpha $ defined by
\begin{equation}
\alpha =\chi -e\Phi /2\pi .  \label{Replace}
\end{equation}%
For the case of a field with periodicity condition (\ref{PerBC}), the
expressions of the renormalized VEVs for the energy density and stresses are
gien by the previous formulas where now the parameter $\alpha $ is defined
as in Eq. (\ref{Replace}).

In general, the fermionic modes in the background of the magnetic vortex are
divided into two classes, regular and irregular (square integrable) ones.
For given $q$ and $\alpha $, the irregular mode corresponds to the value of $%
j$ for which $q|j+\alpha |<1/2$. If we present the parameter $\alpha $ in
the form (\ref{alf}), then the irregular mode is present if $|\alpha
_{0}|>(1-1/q)/2$. This mode corresponds to $j=-n_{0}-$sgn$(\alpha _{0})/2$.
Note that, in a conical space, under the condition $|\alpha _{0}|\leqslant
(1-1/q)/2$, there are no square integrable irregular modes. As we have
already mentioned, there is a one-parameter family of allowed boundary
conditions for irregular modes. These modes are parameterized by the angle $%
\theta $, $0\leqslant \theta <2\pi $ (see Ref. \cite{Sous89}). For $|\alpha
_{0}|<1/2$, the boundary condition, used in deriving mode functions (\ref%
{psi0}), corresponds to $\theta =3\pi /2$. If $\alpha $ is a half-integer,
the irregular mode corresponds to $j=-\alpha $ and for the corresponding
boundary condition one has $\theta =0$. Note that in both cases there are no
bound states.

\section{Energy-momentum tensor inside a circular boundary}

\label{sec:EMTinside}

We turn to the investigation of the effect of a circular boundary on the VEV
of the energy-momentum tensor for a spinor field. We assume that on the
circle the field obeys the MIT bag boundary condition (\ref{BCMIT}). First
we consider the region inside the boundary. In this region the
negative-energy eigenspinors are given by the expression \cite{Beze10b}
\begin{equation}
\psi _{\gamma j}^{(-)}=\varphi _{0}e^{-iqj\phi +iEt}\left(
\begin{array}{c}
\frac{\epsilon _{j}\gamma e^{-iq\phi /2}}{E+m}J_{\beta _{j}+\epsilon
_{j}}(\gamma r) \\
e^{iq\phi /2}J_{\beta _{j}}(\gamma r)%
\end{array}%
\right) ,  \label{psijInt}
\end{equation}%
with the same notations as in Eq. (\ref{psi0}). From the boundary condition
at $r=a$ it follows that the allowed values of $\gamma $ are solutions of
the equation%
\begin{equation}
J_{\beta _{j}}(\gamma a)-\frac{\gamma \epsilon _{j}J_{\beta _{j}+\epsilon
_{j}}(\gamma a)}{m+\sqrt{\gamma ^{2}+m^{2}}}=0.  \label{gamVal}
\end{equation}%
For a given $\beta _{j}$, Eq. (\ref{gamVal}) has an infinite number of
solutions which we denote by $\gamma a=\gamma _{\beta _{j},l}$, $%
l=1,2,\ldots $. The normalization coefficient in Eq. (\ref{psijInt}) is
given by the expression%
\begin{equation}
\varphi _{0}^{2}=\frac{\gamma T_{\beta _{j}}(\gamma a)}{2\phi _{0}a}\frac{m+E%
}{E},  \label{phi0T}
\end{equation}%
with the notation%
\begin{equation}
T_{\beta _{j}}(y)=\frac{y}{J_{\beta _{j}}^{2}(y)}\Big[y^{2}+\left( \mu
-\epsilon _{j}\beta _{j}\right) \left( \mu +\sqrt{y^{2}+\mu ^{2}}\right) -%
\frac{y^{2}}{2\sqrt{y^{2}+\mu ^{2}}}\Big]^{-1},  \label{Tnu}
\end{equation}%
and $\mu =ma$.

Substituting the mode functions (\ref{psijInt}) into Eq. (\ref{modesum})
with $\sum_{\sigma }=\sum_{j}\sum_{l=1}^{\infty }$, for the VEVs of the
separate components we find%
\begin{eqnarray}
\left\langle T_{0}^{0}\right\rangle &=&-\frac{q}{4\pi a}\sum_{j}\sum_{l=1}^{%
\infty }\gamma T_{\beta _{j}}(\gamma a)\left[ (E-m)J_{\beta _{j}+\epsilon
_{j}}^{2}(\gamma r)+(E+m)J_{\beta _{j}}^{2}(\gamma r)\right] ,  \notag \\
\left\langle T_{1}^{1}\right\rangle &=&-\frac{q}{4\pi a}\sum_{j}\sum_{l=1}^{%
\infty }\epsilon _{j}\frac{\gamma ^{3}}{E}T_{\beta _{j}}(\gamma a)[J_{\beta
_{j}}^{\prime }(\gamma r)J_{\beta _{j}+\epsilon _{j}}(\gamma r)-J_{\beta
_{j}+\epsilon _{j}}^{\prime }(\gamma r)J_{\beta _{j}}(\gamma r)],
\label{T11Gen} \\
\left\langle T_{2}^{2}\right\rangle &=&\frac{q}{4\pi a}\sum_{j}\sum_{l=1}^{%
\infty }\frac{\gamma ^{3}}{E}T_{\beta _{j}}(\gamma a)\frac{2\beta
_{j}+\epsilon _{j}}{\gamma r}J_{\beta _{j}}(\gamma r)J_{\beta _{j}+\epsilon
_{j}}(\gamma r),  \notag
\end{eqnarray}%
with $\gamma =\gamma _{\beta _{j},l}/a$. Here we assume that a cutoff
function is introduced without explicitly writing it. The specific form of
this function is not important for the discussion below.

For the summation of the series over $l$ in Eq. (\ref{T11Gen}) we use the
summation formula (see Refs. \cite{Saha04,Saha08Book})%
\begin{eqnarray}
&&\sum_{l=1}^{\infty }f(\gamma _{\beta _{j},l})T_{\beta }(\gamma _{\beta
_{j},l})=\int_{0}^{\infty }dx\,f(x)-\frac{1}{\pi }\int_{0}^{\infty }dx
\notag \\
&&\quad \times \lbrack e^{-\beta _{j}\pi i}f(xe^{\pi i/2})L_{\beta
_{j}}^{(+)}(x)+e^{\beta _{j}\pi i}f(xe^{-\pi i/2})L_{\beta _{j}}^{(+)\ast
}(x)],  \label{SumForm}
\end{eqnarray}%
where
\begin{equation}
L_{\beta _{j}}^{(+)}(x)=\frac{K_{\beta _{j}}^{(+)}(x)}{I_{\beta
_{j}}^{(+)}(x)},  \label{Lbet}
\end{equation}%
and the asterisk means complex conjugate. In Eq. (\ref{Lbet}), for a given
function $F(x)$, we use the notation%
\begin{equation}
F^{(+)}(x)=\left\{
\begin{array}{cc}
xF^{\prime }(x)+(\mu +\sqrt{\mu ^{2}-x^{2}}-\epsilon _{j}\beta _{j})F(x), &
x<\mu , \\
xF^{\prime }(x)+\left( \mu +i\sqrt{x^{2}-\mu ^{2}}-\epsilon _{j}\beta
_{j}\right) F(x), & x\geqslant \mu .%
\end{array}%
\right.  \label{F+Int}
\end{equation}%
Note that for $x<\mu $ one has $F^{(+)\ast }(x)=F^{(+)}(x)$. By using the
properties of the modified Bessel functions, the function $L_{\beta
_{j}}^{(+)}(x)$ can be presented in the form%
\begin{equation}
L_{\beta _{j}}^{(+)}(x)=\frac{W_{\beta _{j},\beta _{j}+\epsilon
_{j}}^{(+)}(x)+i\sqrt{1-\mu ^{2}/x^{2}}}{U_{\beta _{j},\beta _{j}+\epsilon
_{j}}^{(I)}(x)},  \label{KIratio}
\end{equation}%
with the notations defined by
\begin{eqnarray}
W_{\nu ,\sigma }^{(\pm )}(x) &=&x\left[ I_{\nu }(x)K_{\nu }(x)-I_{\sigma
}(x)K_{\sigma }(x)\right]  \notag \\
&&\pm \mu \left[ I_{\sigma }(x)K_{\nu }(x)-I_{\nu }(x)K_{\sigma }(x)\right] ,
\notag \\
U_{\nu ,\sigma }^{(I)}(x) &=&x[I_{\nu }^{2}(x)+I_{\sigma }^{2}(x)]+2\mu
I_{\nu }(x)I_{\sigma }(x).  \label{Wbet}
\end{eqnarray}%
The function $W_{\nu ,\sigma }^{(-)}(x)$ will appear in the expressions for
the VEV in the exterior region (see Sect. \ref{sec:EMToutside}).

Applying to the series over $l$ in Eq. (\ref{T11Gen}) the summation formula,
it can be seen that the terms in the VEVs corresponding to the first
integral in the right-hand side of Eq. (\ref{SumForm}) coincide with the
corresponding VEVs in a boundary-free conical space. As a result, after the
application of formula (\ref{SumForm}), the VEV of the energy-momentum
tensor is presented in the decomposed form
\begin{equation}
\langle T_{k}^{i}\rangle =\langle T_{k}^{i}\rangle _{0,\text{ren}}+\langle
T_{k}^{i}\rangle _{\text{b}},  \label{EMTdecomp}
\end{equation}%
with $\langle T_{k}^{i}\rangle _{\text{b}}$ being the part induced by the
circular boundary. For the functions $f(x)$ corresponding to Eq. (\ref%
{T11Gen}), in the second term on the right-hand side of Eq. (\ref{SumForm}),
the part of the integral over the region $(0,\mu )$ vanishes. As a result,
the boundary-induced contributions in the interior region are given by the
expressions%
\begin{eqnarray}
\langle T_{0}^{0}\rangle _{\text{b}} &=&-\frac{q}{2\pi ^{2}}%
\sum_{j}\int_{m}^{\infty }dx\,\frac{\sqrt{x^{2}-m^{2}}}{U_{\beta _{j},\beta
_{j}+\epsilon _{j}}^{(I)}(ax)}\{m[I_{\beta _{j}}^{2}(rx)+I_{\beta
_{j}+\epsilon _{j}}^{2}(rx)]  \notag \\
&&+x[I_{\beta _{j}}^{2}(rx)-I_{\beta _{j}+\epsilon _{j}}^{2}(rx)]W_{\beta
_{j},\beta _{j}+\epsilon _{j}}^{(+)}(ax)\},  \notag \\
\langle T_{1}^{1}\rangle _{\text{b}} &=&-\frac{q}{2\pi ^{2}}%
\sum_{j}\int_{m}^{\infty }dx\,\frac{x^{3}W_{\beta _{j},\beta _{j}+\epsilon
_{j}}^{(+)}(ax)}{\sqrt{x^{2}-m^{2}}}\frac{I_{\beta _{j}}^{\prime
}(rx)I_{\beta _{j}+\epsilon _{j}}(rx)-I_{\beta _{j}}(rx)I_{\beta
_{j}+\epsilon _{j}}^{\prime }(rx)}{U_{\beta _{j},\beta _{j}+\epsilon
_{j}}^{(I)}(ax)},  \label{TiiInt} \\
\langle T_{2}^{2}\rangle _{\text{b}} &=&\frac{q}{2\pi ^{2}r}\sum_{j}\left(
2\epsilon _{j}\beta _{j}+1\right) \int_{m}^{\infty }dx\,\frac{x^{2}W_{\beta
_{j},\beta _{j}+\epsilon _{j}}^{(+)}(ax)}{\sqrt{x^{2}-m^{2}}}\frac{I_{\beta
_{j}}(rx)I_{\beta _{j}+\epsilon _{j}}(rx)}{U_{\beta _{j},\beta _{j}+\epsilon
_{j}}^{(I)}(ax)}.  \notag
\end{eqnarray}%
For points away from the circular boundary and the cone apex, the
boundary-induced contributions, given by Eq. (\ref{TiiInt}), are finite and
the renormalization is reduced to that for the boundary-free geometry. The
latter we have discussed in the previous section.

Under the change $\alpha \rightarrow -\alpha $, $j\rightarrow -j$, we have $%
\beta _{j}\rightarrow \beta _{j}+\epsilon _{j}$, $\beta _{j}+\epsilon
_{j}\rightarrow \beta _{j}$. From here it follows that, under this change,
the functions $W_{\beta _{j},\beta _{j}+\epsilon _{j}}^{(+)}(ax)$ and $%
U_{\beta _{j},\beta _{j}+\epsilon _{j}}^{(I)}(ax)$ are odd and even
functions, respectively. Now, from Eq. (\ref{TiiInt}) we see that the
boundary-induced parts in the components of the energy-momentum tensor are
even functions of $\alpha $. They are periodic functions of the parameter $%
\alpha $ with the period equal to 1. Consequently, if we present this
parameter in the form (\ref{alf}) with $n_{0}$ being an integer, then the
VEV of the energy-momentum tensor depends on $\alpha _{0}$ alone. Note that,
by using the recurrence relations for the modified Bessel function, the
radial stress can also be written in the form
\begin{eqnarray}
\langle T_{1}^{1}\rangle _{\text{b}} &=&-\frac{q}{2\pi ^{2}}%
\sum_{j}\int_{m}^{\infty }dx\,\frac{x^{3}W_{\beta _{j},\beta _{j}+\epsilon
_{j}}^{(+)}(ax)}{\sqrt{x^{2}-m^{2}}}\,  \notag \\
&&\times \frac{I_{\beta _{j}+\epsilon _{j}}^{2}(rx)-I_{\beta _{j}}^{2}(rx)}{%
U_{\beta _{j},\beta _{j}+\epsilon _{j}}^{(I)}(ax)}-\langle T_{2}^{2}\rangle
_{\text{b}}.  \label{T11b2}
\end{eqnarray}%
Now, it is easy to explicitly check that the trace relation $\langle
T_{i}^{i}\rangle _{\text{b}}=m\langle \bar{\psi}\psi \rangle _{\text{b}}$ is
satisfied.

In the case of a massless field the expressions for the boundary-induced
parts in the VEVs take the form
\begin{eqnarray}
\langle T_{0}^{0}\rangle _{\text{b}} &=&-\frac{q}{2\pi ^{2}a^{3}}%
\sum_{j}\int_{0}^{\infty }dx\,x^{2}V_{\beta _{j},\beta _{j}+\epsilon
_{j}}^{(I)}(x)[I_{\beta _{j}}^{2}(xr/a)-I_{\beta _{j}+\epsilon
_{j}}^{2}(xr/a)],  \notag \\
\langle T_{2}^{2}\rangle _{\text{b}} &=&\frac{q}{2\pi ^{2}a^{2}r}%
\sum_{j}\left( 2\epsilon _{j}\beta _{j}+1\right) \int_{0}^{\infty
}dx\,xV_{\beta _{j},\beta _{j}+\epsilon _{j}}^{(I)}(x)I_{\beta
_{j}}(xr/a)I_{\beta _{j}+\epsilon _{j}}(xr/a),  \label{T22bm0}
\end{eqnarray}%
where we have defined%
\begin{equation}
V_{\nu ,\sigma }^{(I)}(x)=\frac{I_{\nu }(x)K_{\nu }(x)-I_{\sigma
}(x)K_{\sigma }(x)}{I_{\nu }^{2}(x)+I_{\sigma }^{2}(x)},  \label{Vnusig}
\end{equation}%
and for the radial stress we have $\langle T_{1}^{1}\rangle _{\text{b}%
}=-\langle T_{0}^{0}\rangle _{\text{b}}-\langle T_{2}^{2}\rangle _{\text{b}}$%
.

Let us consider asymptotic behavior of the VEV for the energy-momentum
tensor near the cone apex and near the boundary. In the limit $r\rightarrow
0 $ we use the expansion for the modified Bessel functions for small values
of the argument. Writing the parameter $\alpha $ in the form (\ref{alf}), it
is seen that the dominant contribution comes from the term with $%
j=-n_{0}-1/2 $ for $\alpha _{0}>0$ and from the term $j=-n_{0}+1/2$ for $%
\alpha _{0}<0$. The leading terms in the expansions over $r/a$ are given by
the expressions%
\begin{eqnarray}
\langle T_{0}^{0}\rangle _{\text{b}} &\approx &\frac{qa^{-3}}{2^{2q_{\alpha
}}\pi ^{2}}\frac{(r/a)^{2q_{\alpha }-1}}{\Gamma ^{2}(q_{\alpha }+1/2)}%
\int_{\mu }^{\infty }dx\,\frac{x^{2q_{\alpha }}\sqrt{x^{2}-\mu ^{2}}}{%
U_{q_{\alpha }+1/2,q_{\alpha }-1/2}^{(I)}(x)}[W_{q_{\alpha }+1/2,q_{\alpha
}-1/2}^{(+)}(x)-\mu /x],  \notag \\
\langle T_{1}^{1}\rangle _{\text{b}} &\approx &-\frac{qa^{-3}}{2^{2q_{\alpha
}}\pi ^{2}}\frac{(r/a)^{2q_{\alpha }-1}}{(2q_{\alpha }+1)\Gamma
^{2}(q_{\alpha }+1/2)}\int_{\mu }^{\infty }dx\,\frac{x^{2q_{\alpha }+2}}{%
\sqrt{x^{2}-\mu ^{2}}}\frac{W_{q_{\alpha }+1/2,q_{\alpha }-1/2}^{(+)}(x)}{%
U_{q_{\alpha }+1/2,q_{\alpha }-1/2}^{(I)}(x)},  \label{Tikcent}
\end{eqnarray}%
where
\begin{equation}
q_{\alpha }=q(1/2-|\alpha _{0}|).  \label{qalf}
\end{equation}%
For the azimuthal stress one has $\langle T_{2}^{2}\rangle _{\text{b}%
}=2q_{\alpha }\langle T_{1}^{1}\rangle _{\text{b}}$. For $\alpha _{0}=0$ the
dominant contribution comes from the terms $j=-n_{0}\pm 1/2$ and the
corresponding asymptotics are obtained from Eq. (\ref{Tikcent}) taking $%
q_{\alpha }=q/2$ with an additional factor 2. As it is seen from Eq. (\ref%
{Tikcent}), the boundary induced VEVs vanish on the cone apex for $|\alpha
_{0}|<(1-1/q)/2$ and diverge when $|\alpha _{0}|>(1-1/q)/2$. In particular,
the VEVs diverge for a magnetic flux in background of Minkowski spacetime.

For points near the boundary, the dominant contribution to the VEVs come
from large values of $j$. Introducing in Eq. (\ref{TiiInt}) a new
integration variable $x=\beta _{j}y$, we use the uniform asymptotic
expansions for the modified Bessel functions \cite{Abra72}. From these
expansions it follows that to the leading order one has%
\begin{eqnarray}
I_{\beta _{j}}^{2}(\beta _{j}z)-I_{\beta _{j}+\epsilon _{j}}^{2}(\beta
_{j}z) &\sim &\frac{e^{2\beta _{j}\eta (z)}}{\pi \beta _{j}z^{2}}[\epsilon
_{j}-t(z)],  \notag \\
I_{\beta _{j}}^{2}(\beta _{j}z)+I_{\beta _{j}+\epsilon _{j}}^{2}(\beta
_{j}z) &\sim &\frac{e^{2\beta _{j}\eta (z)}}{\pi \beta _{j}z^{2}}%
[1/t(z)-\epsilon _{j}],  \label{Ias}
\end{eqnarray}%
and%
\begin{equation}
K_{\beta _{j}}(\beta _{j}z)I_{\beta _{j}}(\beta _{j}z)-K_{\beta
_{j}+\epsilon _{j}}(\beta _{j}z)\;I_{\beta _{j}+\epsilon _{j}}(\beta
_{j}z)\sim \frac{\epsilon _{j}t^{3}(z)}{2\beta _{j}^{2}},  \label{IKas}
\end{equation}%
with the standard notations $t(z)=1/\sqrt{1+z^{2}}$,%
\begin{equation}
\eta (z)=\sqrt{1+z^{2}}+\ln \left( \frac{z}{1+\sqrt{1+z^{2}}}\right) .
\label{etaz}
\end{equation}%
With the help of these expressions, for the leading terms in the asymptotic
expansions over the distance from the boundary one gets%
\begin{eqnarray}
\langle T_{0}^{0}\rangle _{\text{b}} &\approx &-\frac{1/8+\mu }{16\pi
a(a-r)^{2}},\;  \notag \\
\langle T_{2}^{2}\rangle _{\text{b}} &\approx &\frac{a}{a-r}\langle
T_{1}^{1}\rangle _{\text{b}}\approx \frac{1/8-\mu }{16\pi a(a-r)^{2}}.
\label{Tikbound}
\end{eqnarray}%
As it is seen, near the boundary the energy density is negative, whereas the
signs of the stresses depend on the mass.

Now let us consider the limiting case when $r$ is fixed and the radius of
the circle is large. For a massive field, assuming $ma\gg 1$, we see that
the dominant contribution to the VEVs (\ref{TiiInt}) comes from the region
near the lower limit of the integration. To the leading order we have:%
\begin{eqnarray}
\langle T_{0}^{0}\rangle _{\text{b}} &\approx &-\frac{qm^{3}e^{-2ma}}{16%
\sqrt{\pi }(ma)^{3/2}}\sum_{j}[I_{\beta _{j}}^{2}(rm)+I_{\beta _{j}+\epsilon
_{j}}^{2}(rm)],  \notag \\
\langle T_{1}^{1}\rangle _{\text{b}} &\approx &\frac{qm^{3}e^{-2ma}}{16\sqrt{%
\pi }(ma)^{3/2}}\sum_{j}(2\epsilon _{j}\beta _{j}+1)[I_{\beta _{j}}^{\prime
}(rm)I_{\beta _{j}+\epsilon _{j}}(rm)-I_{\beta _{j}}(rm)I_{\beta
_{j}+\epsilon _{j}}^{\prime }(rm)],  \label{TikRad} \\
\langle T_{2}^{2}\rangle _{\text{b}} &\approx &-\frac{qm^{3}e^{-2ma}}{16%
\sqrt{\pi }(ma)^{3/2}}\sum_{j}\frac{\left( 2\epsilon _{j}\beta _{j}+1\right)
^{2}}{rm}I_{\beta _{j}}(rm)I_{\beta _{j}+\epsilon _{j}}(rm).  \notag
\end{eqnarray}%
In this case the boundary induced VEVs are exponentially suppressed. For a
massless field and for large values of the circle radius, the corresponding
behavior is obtained from Eq. (\ref{Tikcent}) taking $\mu =0$. In this case
the decay of the VEVs is of power-law (no summation): $\langle
T_{i}^{i}\rangle _{\text{b}}\sim 1/a^{2(q_{\alpha }+1)}$.

\begin{figure}[tbph]
\begin{center}
\begin{tabular}{cc}
\epsfig{figure=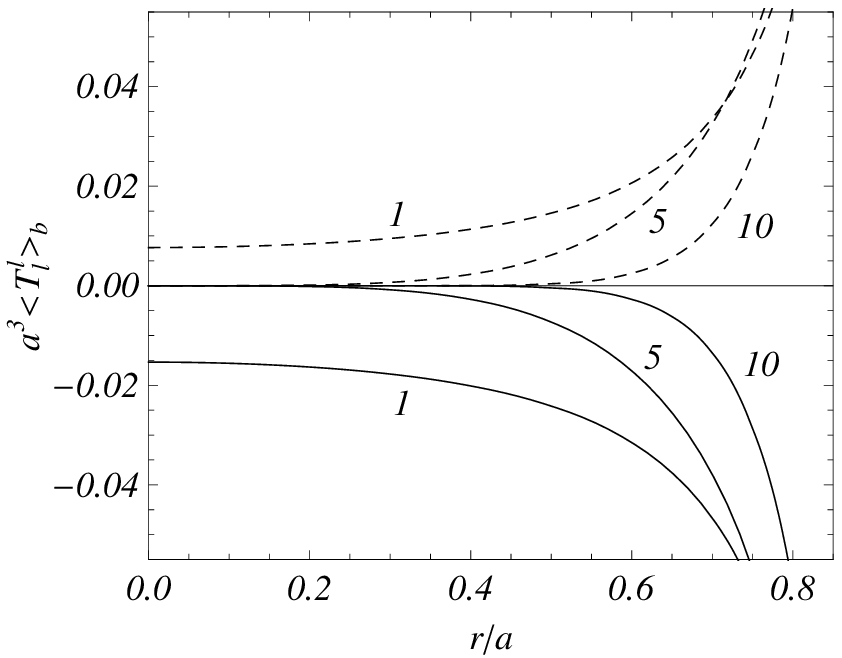,width=7.cm,height=6.cm} & \quad %
\epsfig{figure=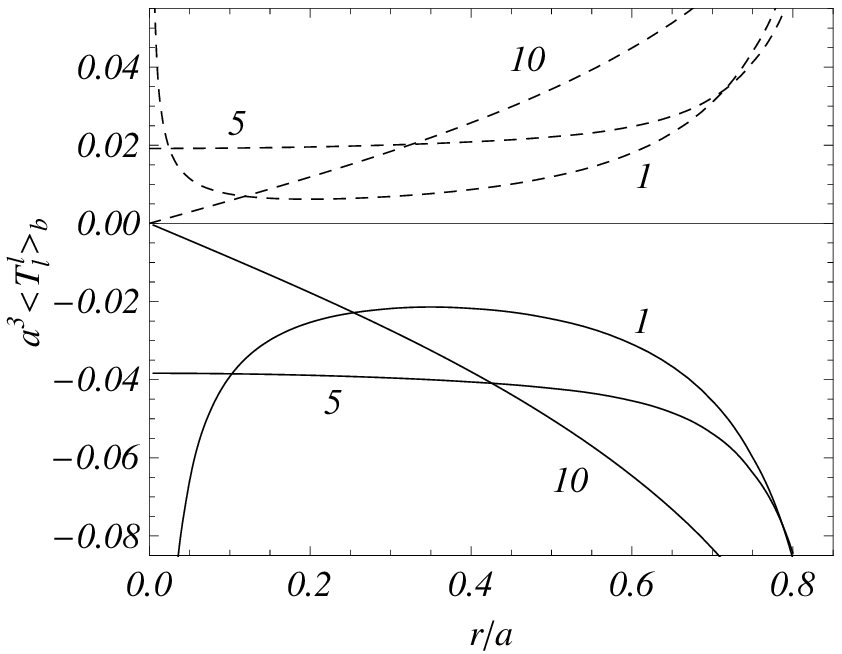,width=7.cm,height=6.cm}%
\end{tabular}%
\end{center}
\caption{Boundary-induced parts in the VEV of the energy density (full
curves) and azimuthal stress (dashed curves) as functions of the radial
coordinate for separate values of the parameter $q$ (numbers near the
curves). The left and right panels are plotted for $\protect\alpha _{0}=0$
and $\protect\alpha _{0}=0.4$, respectively.}
\label{fig2}
\end{figure}

In Fig. \ref{fig2} we display the boundary induced parts in the VEV of the
energy density (full curves) and azimuthal stress (dashed curves) as
functions of the radial coordinate for separate values of the parameter $q$
(numbers near the curves). The left and right panels are plotted for $\alpha
_{0}=0$ and $\alpha _{0}=0.4$, respectively. In the second case, for $q=5,10$
there are no irregular modes and the VEVs are finite on the apex.

\begin{figure}[tbph]
\begin{center}
\begin{tabular}{cc}
\epsfig{figure=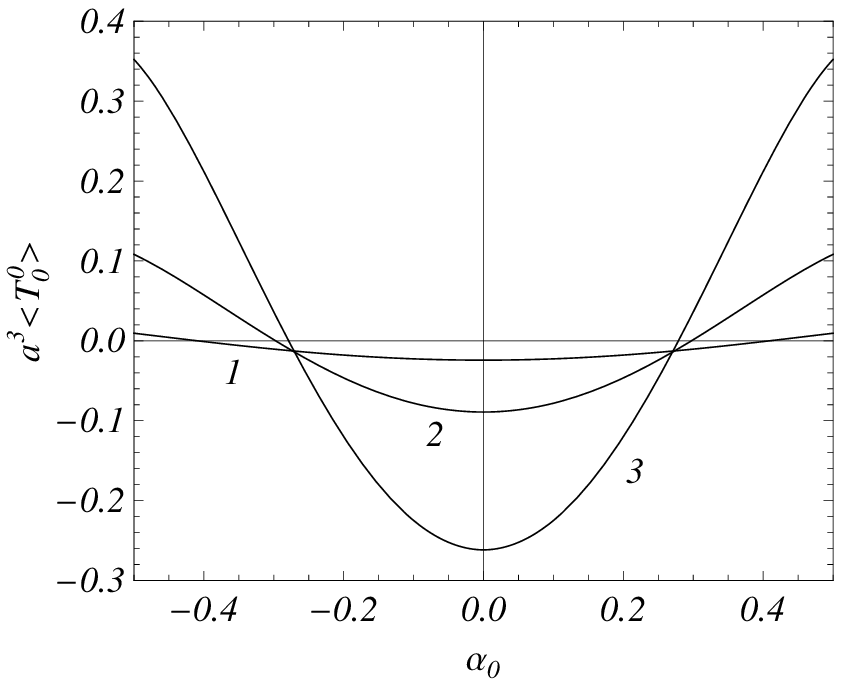,width=7.cm,height=6.cm} & \quad %
\epsfig{figure=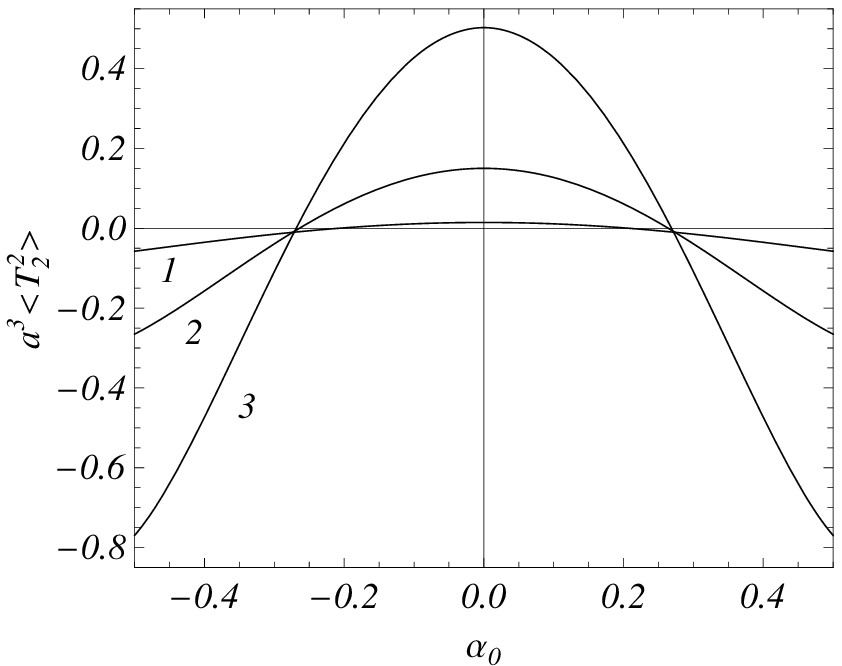,width=7.cm,height=6.cm}%
\end{tabular}%
\end{center}
\caption{Energy density (left panel) and the azimuthal stress (right panel)
for a massless fermionic field as functions of the parameter $\protect\alpha %
_{0}$ for $r/a=0.5$. The numbers near the curves correspond to the values of
the parameter $q$.}
\label{fig3}
\end{figure}

The VEVs of the vacuum energy density and the azimuthal stress for a
massless field are plotted in Fig. \ref{fig3} as functions of the parameter $%
\alpha _{0}$ for fixed value of the radial coordinate corresponding to $%
r/a=0.5$. The numbers near the curves are the values of the parameter $q$.

Various special cases of the general formula (\ref{TiiInt}) can be
considered. In the absence of the magnetic flux one has $\alpha =0$ and the
contributions of the negative and positive values of $j$ to the VEVs
coincide. The corresponding formulas are obtained from Eq. (\ref{TiiInt})
making the replacements%
\begin{equation}
\sum_{j}\rightarrow 2\sum_{j=1/2,3/2,\ldots },\;\beta _{j}\rightarrow
qj-1/2,\;\beta _{j}+\epsilon _{j}\rightarrow qj+1/2.  \label{ReplaceAlf0}
\end{equation}%
In the case $q=1$, we obtain the VEVs induced by the magnetic flux and a
circular boundary in the Minkowski spacetime. And finally, in the simplest
case $\alpha =0$ and $q=1$ one has $\langle T_{i}^{k}\rangle _{0,\text{ren}%
}=0$, and the expressions (\ref{TiiInt}) give the VEVs induced by a circular
boundary in the Minkowski bulk:%
\begin{eqnarray}
\langle T_{0}^{0}\rangle _{\text{b}} &=&-\frac{a^{-3}}{\pi ^{2}}%
\sum_{n=0}^{\infty }\int_{\mu }^{\infty }dx\,\frac{x\sqrt{x^{2}-\mu ^{2}}}{%
U_{n,n+1}^{(I)}(x)}  \notag \\
&&\times \Big\{\left[ I_{n}^{2}(xr/a)-I_{n+1}^{2}(xr/a)\right]
W_{n,n+1}^{(+)}(x)+(\mu /x)\left[ I_{n}^{2}(xr/a)+I_{n+1}^{2}(xr/a)\right] %
\Big\},  \notag \\
\langle T_{1}^{1}\rangle _{\text{b}} &=&-\frac{a^{-3}}{\pi ^{2}}%
\sum_{n=0}^{\infty }\int_{\mu }^{\infty }dx\,\frac{x^{3}W_{n,n+1}^{(+)}(x)}{%
\sqrt{x^{2}-\mu ^{2}}}\frac{I_{n}^{\prime
}(xr/a)I_{n+1}(xr/a)-I_{n+1}^{\prime }(xr/a)I_{n}(xr/a)}{U_{n,n+1}^{(I)}(x)}%
\,,  \label{T11Mink} \\
\langle T_{2}^{2}\rangle _{\text{b}} &=&\frac{a^{-2}}{\pi ^{2}r}%
\sum_{n=0}^{\infty }\left( 2n+1\right) \int_{\mu }^{\infty }dx\,\frac{%
x^{2}W_{n,n+1}^{(+)}(x)}{\sqrt{x^{2}-\mu ^{2}}}\frac{I_{n}(xr/a)I_{n+1}(xr/a)%
}{U_{n,n+1}^{(I)}(x)},  \notag
\end{eqnarray}%
where the functions $W_{n,n+1}^{(+)}(x)$ and $U_{n,n+1}^{(I)}(x)$ are
defined by Eq. (\ref{KIratio}).

\section{VEV outside a circular boundary}

\label{sec:EMToutside}

In this section we consider the region outside a circular boundary with
radius $a$. The corresponding negative-energy mode functions, obeying the
boundary condition (\ref{BCMIT}), are given by the expression \cite{Beze10b}%
\begin{equation}
\psi _{\gamma j}^{(-)}(x)=c_{0}e^{-iqj\phi +iEt}\left(
\begin{array}{c}
\frac{\gamma \epsilon _{j}e^{-iq\phi /2}}{E+m}g_{\beta _{j},\beta
_{j}+\epsilon _{j}}(\gamma a,\gamma r) \\
g_{\beta _{j},\beta _{j}}(\gamma a,\gamma r)e^{iq\phi /2}%
\end{array}%
\right) ,  \label{psiminext}
\end{equation}%
with the function%
\begin{equation}
g_{\nu ,\rho }(x,y)=\bar{Y}_{\nu }^{(-)}(x)J_{\rho }(y)-\bar{J}_{\nu
}^{(-)}(x)Y_{\rho }(y),  \label{gsig}
\end{equation}%
and with $Y_{\nu }(x)$\ being the Neumann function. The barred notation is
defined by the relation%
\begin{equation}
\bar{F}_{\beta _{j}}^{(-)}(z)=-\epsilon _{j}zF_{\beta _{j}+\epsilon
_{j}}(z)-(\sqrt{z^{2}+\mu ^{2}}+\mu )F_{\beta _{j}}(z),  \label{barnot3}
\end{equation}%
with $F=J,Y$ and, as before, $\mu =ma$. For the normalization coefficient in
Eq. (\ref{psiminext}) one has
\begin{equation}
c_{0}^{2}=\frac{2E\gamma }{\phi _{0}(E+m)}[\bar{J}_{\beta
_{j}}^{(-)2}(\gamma a)+\bar{Y}_{\beta _{j}}^{(-)2}(\gamma a)]^{-1}.
\label{c0}
\end{equation}%
The positive-energy eigenspinors are obtained by making use of the relation $%
\psi _{\gamma n}^{(+)}=\sigma _{1}\psi _{\gamma n}^{(-)\ast }$. Note that in
the exterior region the conical singularity is excluded by the boundary and
all modes described by eigenspinors (\ref{psiminext}) are regular.

Substituting the mode functions into the mode-sum formula, the VEVs\ for
separate components of the energy-momentum tensor are written in the form%
\begin{eqnarray}
\left\langle T_{0}^{0}\right\rangle &=&-\frac{q}{4\pi }\sum_{j}\int_{0}^{%
\infty }d\gamma \gamma \frac{(E-m)g_{\beta _{j},\beta _{j}+\epsilon
_{j}}^{2}(\gamma a,\gamma r)+(E+m)g_{\beta _{j},\beta _{j}}^{2}(\gamma
a,\gamma r)}{\bar{J}_{\beta _{j}}^{(-)2}(\gamma a)+\bar{Y}_{\beta
_{j}}^{(-)2}(\gamma a)},  \notag \\
\left\langle T_{1}^{1}\right\rangle &=&\frac{q}{4\pi }\sum_{j}\int_{0}^{%
\infty }d\gamma \frac{\gamma ^{3}}{E}\frac{g_{\beta _{j},\beta _{j}+\epsilon
_{j}}^{2}(\gamma a,\gamma r)+g_{\beta _{j},\beta _{j}}^{2}(\gamma a,\gamma r)%
}{\bar{J}_{\beta _{j}}^{(-)2}(\gamma a)+\bar{Y}_{\beta _{j}}^{(-)2}(\gamma a)%
}-\left\langle T_{2}^{2}\right\rangle ,  \label{T11unReg} \\
\left\langle T_{2}^{2}\right\rangle &=&\frac{q}{4\pi r}\sum_{j}\int_{0}^{%
\infty }d\gamma \frac{(2\beta _{j}+\epsilon _{j})\gamma ^{2}/E}{\bar{J}%
_{\beta _{j}}^{(-)2}(\gamma a)+\bar{Y}_{\beta _{j}}^{(-)2}(\gamma a)}%
g_{\beta _{j},\beta _{j}}(\gamma a,\gamma r)g_{\beta _{j},\beta
_{j}+\epsilon _{j}}(\gamma a,\gamma r).  \notag
\end{eqnarray}%
As before, we assume the presence of a cutoff function which makes the
expression on the right-hand sides of Eq. (\ref{T11unReg}) finite. Similar
to the interior region, the VEVs\ outside a circular boundary may be written
in the decomposed form (\ref{EMTdecomp}).

In order to find an explicit expression for the boundary-induced part, we
note that the boundary-free part is given by Eq. (\ref{Tii0}). For the
evaluation of the difference between the total VEV and the boundary-free
part, we use the identities%
\begin{eqnarray}
\frac{g_{\beta _{j},\lambda }^{2}(x,y)}{\bar{J}_{\beta _{j}}^{(-)2}(x)+\bar{Y%
}_{\beta _{j}}^{(-)2}(x)} &=&J_{\lambda }^{2}(y)-\frac{1}{2}\sum_{l=1,2}%
\frac{\bar{J}_{\beta _{j}}^{(-)}(x)}{\bar{H}_{\beta _{j}}^{(-,l)}(x)}%
H_{\lambda }^{(l)2}(y),  \notag \\
\frac{g_{\beta _{j},\lambda }(x,y)g_{\beta _{j},\lambda +\epsilon _{j}}(x,y)%
}{\bar{J}_{\beta _{j}}^{(-)2}(x)+\bar{Y}_{\beta _{j}}^{(-)2}(x)}
&=&J_{\lambda }(y)J_{\lambda +\epsilon _{j}}(y)-\frac{1}{2}\sum_{l=1,2}\frac{%
\bar{J}_{\beta _{j}}^{(-)}(x)}{\bar{H}_{\beta _{j}}^{(-,l)}(x)}H_{\lambda
}^{(l)}(y)H_{\lambda +\epsilon _{j}}^{(l)}(y),  \label{Ident1}
\end{eqnarray}%
with $\lambda =\beta _{j},\beta _{j}+\epsilon _{j}$, and with $H_{\nu
}^{(l)}(x)$ being the Hankel function.

In this way, for the boundary induced parts we find the expressions%
\begin{eqnarray}
\langle T_{0}^{0}\rangle _{\text{b}} &=&\frac{q}{8\pi }\sum_{j}\sum_{l=1,2}%
\int_{0}^{\infty }d\gamma \gamma \frac{\bar{J}_{\beta _{j}}^{(-)}(\gamma a)}{%
\bar{H}_{\beta _{j}}^{(-,l)}(\gamma ax)}\left[ (E-m)H_{\beta _{j}+\epsilon
_{j}}^{(l)2}(\gamma r)+(E+m)H_{\beta _{j}}^{(l)2}(\gamma r)\right] ,  \notag
\\
\langle T_{1}^{1}\rangle _{\text{b}} &=&-\frac{q}{8\pi }\sum_{j}\sum_{l=1,2}%
\int_{0}^{\infty }d\gamma \frac{\gamma ^{3}}{E}\frac{\bar{J}_{\beta
_{j}}^{(-)}(\gamma a)}{\bar{H}_{\beta _{j}}^{(-,l)}(\gamma a)}\left[
H_{\beta _{j}+\epsilon _{j}}^{(l)2}(\gamma r)+H_{\beta _{j}}^{(l)2}(\gamma r)%
\right] -\langle T_{2}^{2}\rangle _{\text{b}},  \label{T11bext0} \\
\langle T_{2}^{2}\rangle _{\text{b}} &=&-\frac{q}{8\pi r}\sum_{j}(2\beta
_{j}+\epsilon _{j})\sum_{l=1,2}\int_{0}^{\infty }d\gamma \frac{\gamma ^{2}}{E%
}\frac{\bar{J}_{\beta _{j}}^{(-)}(\gamma a)}{\bar{H}_{\beta
_{j}}^{(-,l)}(\gamma a)}H_{\beta _{j}}^{(l)}(\gamma r)H_{\beta _{j}+\epsilon
_{j}}^{(l)}(\gamma r).  \notag
\end{eqnarray}%
In the complex plane $\gamma $, the integrand of the term with $l=1$ ($l=2$)
decays exponentially in the limit ${\mathrm{Im}}(\gamma )\rightarrow \infty $
[${\mathrm{Im}}(\gamma )\rightarrow -\infty $] for $r>a$ . By using these
properties, we rotate the integration contour in the complex plane $\gamma $
by the angle $\pi /2$ for the term with $l=1$ and by the angle $-\pi /2$ for
the term with $l=2$. The integrals over the segments $(0,im)$ and $(0,-im)$
of the imaginary axis cancel each other. Introducing the modified Bessel
functions, the boundary-induced parts are presented in the form%
\begin{eqnarray}
\langle T_{0}^{0}\rangle _{\text{b}} &=&-\frac{q}{2\pi ^{2}}%
\sum_{j}\int_{m}^{\infty }dx\,x\frac{\sqrt{x^{2}-m^{2}}}{U_{\beta _{j},\beta
_{j}+\epsilon _{j}}^{(K)}(ax)}\left\{ m[K_{\beta _{j}+\epsilon
_{j}}^{2}(xr)+K_{\beta _{j}}^{2}(xr)]\right.  \notag \\
&&\left. +x[K_{\beta _{j}}^{2}(xr)-K_{\beta _{j}+\epsilon
_{j}}^{2}(xr)]W_{\beta _{j},\beta _{j}+\epsilon _{j}}^{(-)}(ax)\right\} ,
\notag \\
\langle T_{1}^{1}\rangle _{\text{b}} &=&-\frac{q}{2\pi ^{2}}%
\sum_{j}\int_{m}^{\infty }dx\,x^{3}\frac{K_{\beta _{j}+\epsilon
_{j}}^{2}(xr)-K_{\beta _{j}}^{2}(xr)}{\sqrt{x^{2}-m^{2}}}\frac{W_{\beta
_{j},\beta _{j}+\epsilon _{j}}^{(-)}(ax)}{U_{\beta _{j},\beta _{j}+\epsilon
_{j}}^{(K)}(ax)}-\langle T_{2}^{2}\rangle _{\text{b}},  \label{T11bext} \\
\langle T_{2}^{2}\rangle _{\text{b}} &=&-\frac{q}{2\pi ^{2}r}\sum_{j}(2\beta
_{j}\epsilon _{j}+1)\int_{m}^{\infty }dx\,x^{2}\frac{K_{\beta
_{j}}(xr)K_{\beta _{j}+\epsilon _{j}}(xr)}{\sqrt{x^{2}-m^{2}}}\frac{W_{\beta
_{j},\beta _{j}+\epsilon _{j}}^{(-)}(ax)}{U_{\beta _{j},\beta _{j}+\epsilon
_{j}}^{(K)}(ax)}.  \notag
\end{eqnarray}%
Here we have introduced the notation%
\begin{equation}
U_{\nu ,\sigma }^{(K)}(x)=x[K_{\nu }^{2}(x)+K_{\sigma }^{2}(x)]+2\mu K_{\nu
}(x)K_{\sigma }(x),  \label{UK}
\end{equation}%
and the notation $W_{\beta _{j},\beta _{j}+\epsilon _{j}}^{(-)}(x)$ is
defined by Eq. (\ref{Wbet}). By taking into account that under the change $%
\alpha \rightarrow -\alpha $, $j\rightarrow -j$, one has $\beta
_{j}\rightarrow \beta _{j}+\epsilon _{j}$, $\beta _{j}+\epsilon
_{j}\rightarrow \beta _{j}$, we conclude that $W_{\beta _{j},\beta
_{j}+\epsilon _{j}}^{(-)}(x)$ and $U_{\beta _{j},\beta _{j}+\epsilon
_{j}}^{(K)}(x)$ are odd and even functions under this change. Now, from Eq. (%
\ref{T11bext}) it follows that the boundary-induced parts are even functions
of $\alpha $. They are periodic with the period equal to~1.

For a massless field the expressions for the boundary-induced parts in the
VEVs simplify to%
\begin{eqnarray}
\langle T_{0}^{0}\rangle _{\text{b}} &=&-\frac{q}{2\pi ^{2}a^{3}}%
\sum_{j}\int_{0}^{\infty }dx\,x^{2}V_{\beta _{j},\beta _{j}+\epsilon
_{j}}^{(K)}(x)[K_{\beta _{j}}^{2}(xr/a)-K_{\beta _{j}+\epsilon
_{j}}^{2}(xr/a)],  \notag \\
\langle T_{2}^{2}\rangle _{\text{b}} &=&-\frac{q}{2\pi ^{2}a^{2}r}%
\sum_{j}(2\beta _{j}\epsilon _{j}+1)\int_{0}^{\infty }dx\,xV_{\beta
_{j},\beta _{j}+\epsilon _{j}}^{(K)}(x)K_{\beta _{j}}(xr/a)K_{\beta
_{j}+\epsilon _{j}}(xr/a),  \label{T22bm0ext}
\end{eqnarray}%
with the notation%
\begin{equation}
V_{\nu ,\sigma }^{(K)}(x)=\frac{I_{\nu }(x)K_{\nu }(x)-I_{\sigma
}(x)K_{\sigma }(x)}{K_{\nu }^{2}(x)+K_{\sigma }^{2}(x)}.  \label{VnusigK}
\end{equation}%
For the radial stress one has $\langle T_{1}^{1}\rangle _{\text{b}}=-\langle
T_{0}^{0}\rangle _{\text{b}}-\langle T_{2}^{2}\rangle _{\text{b}}$. In
particular, for the circle in the Minkowski bulk the corresponding formulas
are obtained from Eq. (\ref{T11Mink}) by the interchange $I\rightleftarrows
K $, replacing $W_{n,n+1}^{(+)}(x)\rightarrow W_{n,n+1}^{(-)}(x)$.

Now we turn to the investigation of the VEVs in the asymptotic regions for
the parameters. First we consider the limit $a\rightarrow 0$ for a fixed
value of $r$. By using the asymptotic formulas for the modified Bessel
functions for small arguments, we can see that the dominant contribution
comes from the term with $j=1/2$ for $-1/2<\alpha _{0}<0$ and from the term $%
j=1/2$ for $0<\alpha _{0}<1/2$. For a massive field to the leading order we
get (no summation over $i$)%
\begin{equation}
\langle T_{i}^{i}\rangle _{\text{b}}\approx -\frac{qm}{\pi ^{2}r^{2}}\frac{%
(a/2r)^{2q_{\alpha }}}{\Gamma ^{2}(q_{\alpha }+1/2)}\int_{mr}^{\infty }dx\,%
\frac{x^{2q_{\alpha }+2}Z_{i}(x)}{\sqrt{x^{2}-m^{2}r^{2}}},  \label{TiiExta0}
\end{equation}%
with the notations%
\begin{eqnarray}
Z_{0}(x) &=&\left( 1-m^{2}r^{2}/x^{2}\right) K_{q_{\alpha }-1/2}^{2}(x),
\notag \\
Z_{1}(x) &=&K_{q_{\alpha }+1/2}^{2}(x)-K_{q_{\alpha }-1/2}^{2}(x)-Z_{2}(x),
\label{Z0} \\
Z_{2}(x) &=&\frac{2q_{\alpha }}{x}K_{q_{\alpha }-1/2}(x)K_{q_{\alpha
}+1/2}(x),  \notag
\end{eqnarray}%
and $q_{\alpha }$ is defined by Eq. (\ref{qalf}). We see that the
boundary-induced part vanishes in the limit $a\rightarrow 0$ for $|\alpha
_{0}|<1/2$. For a massless field the leading term in the corresponding
asymptotic expansion for the azimuthal stress is given by the expression%
\begin{equation}
\langle T_{2}^{2}\rangle _{\text{b}}=-\frac{q}{4\pi r^{3}}\frac{%
(a/2r)^{2q_{\alpha }+1}}{q_{\alpha }^{2}-1/4}\frac{q_{\alpha }\Gamma
(q_{\alpha }+1)\Gamma \left( 2q_{\alpha }+3/2\right) }{\Gamma (q_{\alpha
}+3/2)\Gamma ^{2}(q_{\alpha }+1/2)},  \label{T22Exta0}
\end{equation}%
For the energy density and the radial stress one has the relations%
\begin{equation}
\langle T_{0}^{0}\rangle _{\text{b}}\approx -\frac{q_{\alpha }+1}{q_{\alpha
}+3/2}\langle T_{2}^{2}\rangle _{\text{b}},\;\langle T_{1}^{1}\rangle _{%
\text{b}}\approx \frac{-\langle T_{2}^{2}\rangle _{\text{b}}}{2\left(
q_{\alpha }+3/2\right) }.  \label{T00Exa0}
\end{equation}%
For a massless field the boundary-induced VEVs are suppressed with an
additional factor $a/r$ with respect to the case of a massive field.

Let us consider the limit of large distances from the circle. For a massive
field, assuming $mr\gg 1$, we see that the dominant contribution to the
integrals in Eq. (\ref{T11bext}) comes from the lower limit of the
integration. In the leading order we obtain:
\begin{eqnarray}
\langle T_{0}^{0}\rangle _{\text{b}} &\approx &-\frac{qm^{3}e^{-2mr}}{8\sqrt{%
\pi }(mr)^{5/2}}\sum_{j}\frac{1}{U_{\beta _{j},\beta _{j}+\epsilon
_{j}}^{(K)}(am)},  \notag \\
\langle T_{2}^{2}\rangle _{\text{b}} &\approx &-\frac{qm^{3}e^{-2mr}}{8\sqrt{%
\pi }(mr)^{5/2}}\sum_{j}\frac{(2\beta _{j}\epsilon _{j}+1)W_{\beta
_{j},\beta _{j}+\epsilon _{j}}^{(-)}(am)}{U_{\beta _{j},\beta _{j}+\epsilon
_{j}}^{(K)}(am)},  \label{T22Extlr}
\end{eqnarray}%
and for the radial stress one has $\langle T_{1}^{1}\rangle _{\text{b}}=-$ $%
\langle T_{2}^{2}\rangle _{\text{b}}/(2mr)$. As we could expect, in this
limit the VEVs are exponentially suppressed. The radial stress contains an
additional suppression factor $(mr)^{-1}$. For a massless field the the
leading terms for $r\gg a$ are given by Eqs. (\ref{T22Exta0}) and (\ref%
{T00Exa0}).

It remains to consider the behavior of the VEVs near the boundary. In this
region the dominant contribution comes from large values of $|j|$ and, in
the way similar to that for the interior region, we find%
\begin{eqnarray}
\langle T_{0}^{0}\rangle _{\text{b}} &\approx &\frac{1/8-\mu }{16\pi
a(r-a)^{2}},  \notag \\
\langle T_{2}^{2}\rangle _{\text{b}} &\approx &\frac{a}{a-r}\langle
T_{1}^{1}\rangle _{\text{b}}\approx -\frac{1/8+\mu }{16\pi a(r-a)^{2}}.
\label{T00Extnb}
\end{eqnarray}%
Comparing with Eq. (\ref{Tikbound}), we see that for a massless field the
energy density and the azimuthal stress have opposite signs for the exterior
and interior regions, whereas the radial stress has the same sign.

The boundary-induced parts in the exterior region are displayed in Fig. \ref%
{fig4} as functions of the radial coordinate. The full and dashed curves are
for the energy density and the azimuthal stress, respectively, and the
numbers near the curves correspond to the values of $q$. The left and right
panels are plotted for $\alpha _{0}=0$ and $\alpha _{0}=0.4$, respectively.

\begin{figure}[tbph]
\begin{center}
\begin{tabular}{cc}
\epsfig{figure=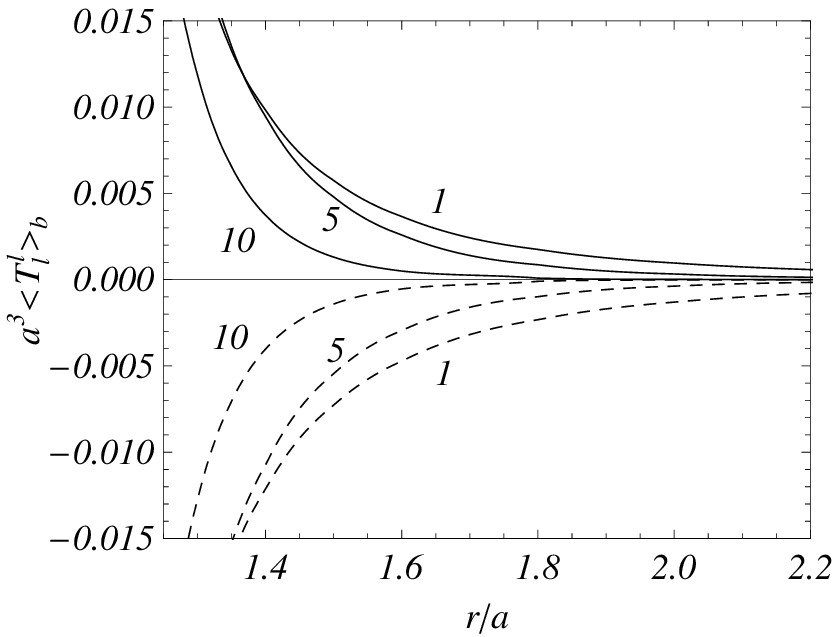,width=7.cm,height=6.cm} & \quad %
\epsfig{figure=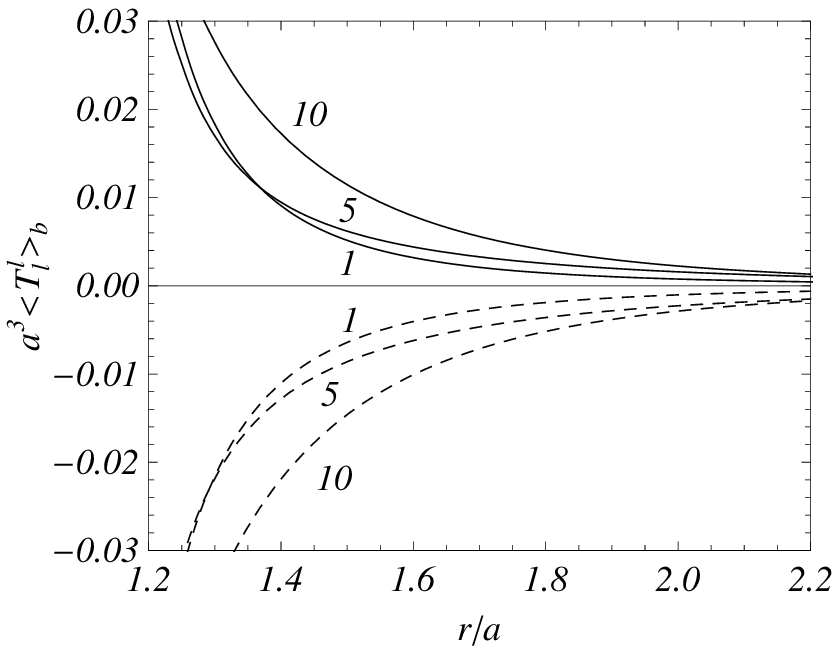,width=7.cm,height=6.cm}%
\end{tabular}%
\end{center}
\caption{The same as in Fig. \protect\ref{fig2} for the region outside a
circular boundary.}
\label{fig4}
\end{figure}

\begin{figure}[tbph]
\begin{center}
\begin{tabular}{cc}
\epsfig{figure=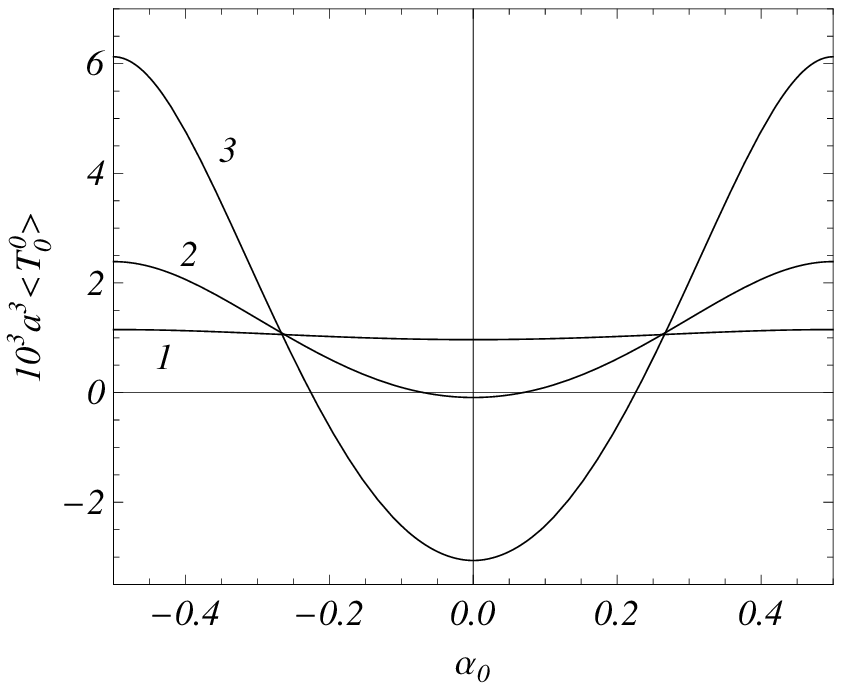,width=7.cm,height=6.cm} & \quad %
\epsfig{figure=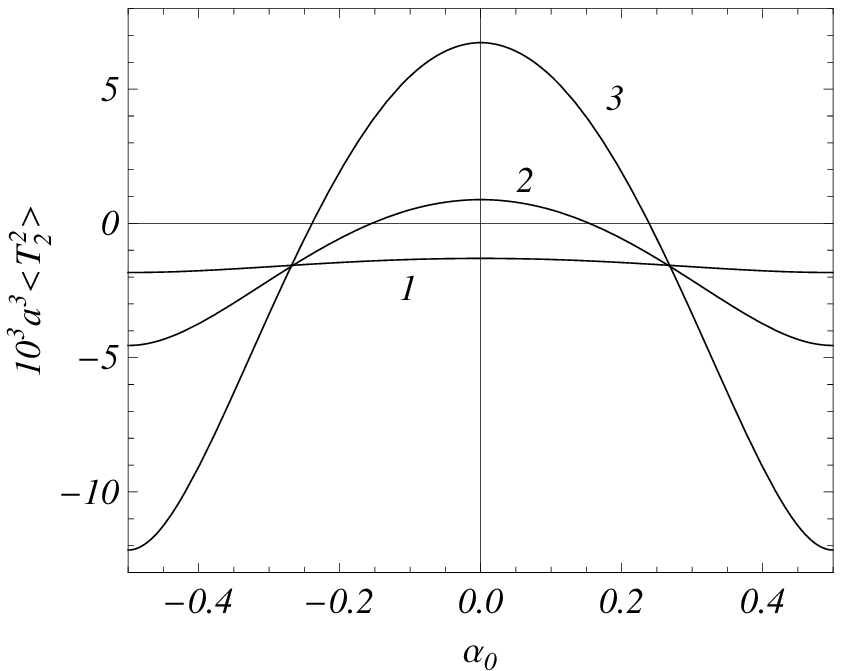,width=7.cm,height=6.cm}%
\end{tabular}%
\end{center}
\caption{The same as in Fig. \protect\ref{fig3} for the region outside a
circular boundary with $r/a=2$.}
\label{fig5}
\end{figure}

In Fig. \ref{fig5} we plot the VEVs of the vacuum energy density and the
azimuthal stress for a massless field versus $\alpha _{0}$ for $r/a=2$. The
numbers near the curves correspond to the values of the parameter $q$.

The results given above can be applied to graphitic cones within
the framework of long-wavelength Dirac-like model for electronic
states in graphene (for a review see \cite{Cast09}). Graphene made
structures have attracted much attention recently due to the
experimental observation of a number of novel electronic
properties. The electronic band structure of graphene close to the
Dirac
points shows a conical dispersion $E(\mathbf{k})=v_{F}|\mathbf{k}|$, where $%
\mathbf{k}$ is the momentum measured relatively to the Dirac points and $%
v_{F}\approx 10^{8}$ cm/s represents the Fermi velocity which
plays the role of a speed of light. The low-energy excitations can
be described by a pair of two-component spinors, $\psi _{J}$,
$J=1,2$, corresponding to the two inequivalent Fermi points of the
Brillouin zone. The components of the spinor $\psi _{J}$
correspond to two triangular sublattices of the honeycomb lattice
of graphene. For a flat graphene sheet, in the absence of
interactions that mix the inequivalent Fermi points, the
electronic states attached to these points will be
independent. The Dirac equation for the corresponding spinors has the form%
\begin{equation}
(iv_{F}^{-1}\gamma ^{0}D_{0}+i\gamma ^{l}D_{l}-m)\psi _{J}=0,\
\label{DeqGraph}
\end{equation}%
where $l=1,2$, and $D_{\mu }=\nabla _{\mu }+ieA_{\mu }$ with $e=-|e|$ for
electrons. The mass (gap) term in (\ref{DeqGraph}) is essential in many
physical application. This gap can be generated by a number of mechanisms.
In particular, they include the breaking of symmetry between two sublattices
by introducing a staggered onsite energy \cite{Seme84} and the deformations
of bonds in the graphene lattice \cite{Cham00}. Another approach is to
attach a graphene monolayer to a substrate the interaction with which breaks
the sublattice symmetry \cite{Giov07}.  Graphitic cones are obtained from
the graphene sheet if one or more sectors with the angle $\pi /6$ are
removed. The opening angle of the cone is related to the number of sectors
removed, $N_{c}$, by the formula $2\pi (1-N_{c}/6)$, with $N_{c}=1,2,\ldots
,5$ (for the electronic properties of graphitic cones see, e.g., \cite%
{Lamm00} and references therein). All these angles have been
observed in experiments \cite{Kris97}. For even values of $N_{c}$
the periodicity conditions do not mix the spinors $\psi _{1}$ and
$\psi _{2}$. The corresponding expressions for the Casimir
densities for finite radius graphitic nanocones are obtained from
the formulas given above with additional factor 2 which takes into
account the presence of two inequivalent Fermi points. In standard
units, the factor $\hbar v_{F}$ appears as well. For odd values of
$N_{c}$ the periodicity condition mixes the spinors corresponding
to inequivalent Fermi points. In this case it is convenient to
combine two spinors $\psi _{1}$ and $\psi _{2}$ in a single
bispinor. The evaluation for the corresponding Casimir densities
can be done in a way similar to that described before.

\section{Half-integer values of the parameter $\protect\alpha $}

\label{sec:EMTspecial}

In this section we consider the VEV of the energy-momentum tensor for
half-integer values of the parameter $\alpha $. In this case the mode with $%
j=-\alpha $ must be considered separately.

\subsection{Boundary-free part}

For half-integer values of $\alpha $, in the boundary-free geometry the
eigenspinors with $j\neq -\alpha $ are still given by Eq. (\ref{psi0}). For
the mode function corresponding to the special mode with $j=-\alpha $ one
has \cite{Beze10b}%
\begin{equation}
\psi _{(0)\gamma ,-\alpha }^{(-)}(x)=\left( \frac{E+m}{\pi \phi _{0}rE}%
\right) ^{1/2}e^{iq\alpha \phi +iEt}\left(
\begin{array}{c}
\frac{\gamma e^{-iq\phi /2}}{E+m}\sin (\gamma r-\gamma _{0}) \\
e^{iq\phi /2}\cos (\gamma r-\gamma _{0})%
\end{array}%
\right) ,  \label{psibetSp}
\end{equation}%
where, as before, $E=\sqrt{\gamma ^{2}+m^{2}}$ and we have defined%
\begin{equation}
\gamma _{0}=\arccos [\sqrt{(E-m)/2E}].  \label{gam0}
\end{equation}%
As it has been noted above, for half-integer values of $\alpha $ the mode
with $j=-\alpha $ corresponds to the irregular mode. The contribution of the
modes with $j\neq -\alpha $ to the VEV of the energy-momentum tensor remains
the same as before. Special consideration is needed for the mode with $%
j=-\alpha $ only. For the contribution of this mode to the VEVs of the
energy density and the radial stress we have the expressions%
\begin{eqnarray}
\left\langle T_{0}^{0}\right\rangle _{0}^{(j=-\alpha )} &=&-\frac{q}{2\pi
^{2}r}\int_{0}^{\infty }d\gamma \left[ E+m\cos (2\gamma r-2\gamma _{0})%
\right] ,  \notag \\
\left\langle T_{1}^{1}\right\rangle _{0}^{(j=-\alpha )} &=&\frac{q}{2\pi
^{2}r}\int_{0}^{\infty }d\gamma \left( E-m^{2}/E\right) ,  \label{T00Ir}
\end{eqnarray}%
and the contribution to the azimuthal stress vanishes. As we have done in
Sect. \ref{sec:BoundFree}, for the regularization of the expressions (\ref%
{T00Ir}) we introduce the cutoff function $e^{-s\gamma ^{2}}$. After the
integration we find the following expressions%
\begin{eqnarray}
\left\langle T_{0}^{0}\right\rangle _{0,\text{reg}}^{(j=-\alpha )} &=&-\frac{%
qm^{2}}{8\pi ^{2}r}\Big\{e^{sm^{2}/2}\left[ K_{0}(sm^{2}/2)+K_{1}(sm^{2}/2)%
\right]  \notag \\
&&+4K_{1}(2mr)-4K_{0}(2mr)+o(s)\Big\},  \notag \\
\left\langle T_{1}^{1}\right\rangle _{0,\text{reg}}^{(j=-\alpha )} &=&-\frac{%
qm^{2}}{8\pi ^{2}r}e^{sm^{2}/2}\left[ K_{0}(sm^{2}/2)-K_{1}(sm^{2}/2)\right]
+o(s).  \label{T11Irreg}
\end{eqnarray}

In order to obtain the total VEV we should add the regularized part
corresponding to the modes with $j\neq -\alpha $. For half-integer values of
$\alpha $, for the series in the contribution of these modes one has
\begin{equation}
\sum_{j\neq -\alpha }I_{\beta _{j}}(x)=\sum_{j\neq -\alpha }I_{\beta
_{j}+\epsilon _{j}}(x)=\sum_{n=1}^{\infty }\left[ I_{qn-1/2}(x)+I_{qn+1/2}(x)%
\right] .  \label{SumIbetHI}
\end{equation}%
As a result, for this part in the regularized VEV of the radial stress one
finds%
\begin{equation}
\langle T_{1}^{1}\rangle _{0,\text{reg}}^{(j\neq -\alpha )}=\frac{qe^{m^{2}s}%
}{\left( 2\pi \right) ^{3/2}}\int_{0}^{1/(2s)}dy\frac{%
y^{1/2}e^{-m^{2}/(2y)-r^{2}y}}{\sqrt{1-2ys}}\sum_{n=1}^{\infty }\left[
I_{qn-1/2}(r^{2}y)+I_{qn+1/2}(r^{2}y)\right] .  \label{T11Rreg}
\end{equation}%
The corresponding parts in the energy density and the azimuthal stress are
given by the relations%
\begin{eqnarray}
\langle T_{0}^{0}\rangle _{0,\text{reg}}^{(j\neq -\alpha )} &=&-\left(
2+r\partial _{r}\right) \langle T_{1}^{1}\rangle _{0,\text{reg}}^{(j\neq
-\alpha )}+m\langle \bar{\psi}\psi \rangle _{0,\text{reg}}^{(j\neq -\alpha
)},  \notag \\
\langle T_{2}^{2}\rangle _{0,\text{reg}}^{(j\neq -\alpha )} &=&\left(
1+r\partial _{r}\right) \langle T_{1}^{1}\rangle _{0,\text{reg}}^{(j\neq
-\alpha )},  \label{T00Rreg}
\end{eqnarray}%
where%
\begin{eqnarray}
\langle \bar{\psi}\psi \rangle _{0,\text{reg}}^{(j\neq -\alpha )} &=&-\frac{%
qme^{m^{2}s}}{(2\pi )^{3/2}}\int_{0}^{r^{2}/2s}dy\frac{%
y^{-1/2}e^{-m^{2}/2y-r^{2}y}}{\sqrt{1-2ys}}  \notag \\
&&\times \sum_{n=1}^{\infty }\left[ I_{qn-1/2}(r^{2}y)+I_{qn+1/2}(r^{2}y)%
\right] .  \label{FCRreg}
\end{eqnarray}%
The fermionic condensate has been considered in Ref. \cite{Bell11}, and here
we need to consider the radial stress only. After the summation over $n$ by
using the formula given in Sect. \ref{sec:BoundFree}, we find the following
representation%
\begin{eqnarray}
\langle T_{1}^{1}\rangle _{0,\text{reg}}^{(j\neq -\alpha )} &=&\frac{%
e^{m^{2}s}}{(2\pi )^{3/2}}\int_{0}^{1/2s}dy\frac{y^{1/2}e^{-m^{2}/2y}}{\sqrt{%
1-2ys}}+\frac{qm^{2}}{8\pi ^{2}r}e^{sm^{2}/2}\left[
K_{0}(sm^{2}/2)-K_{1}(sm^{2}/2)\right]  \notag \\
&&+\frac{m^{3}}{\pi }e^{m^{2}s}\bigg[\sum_{l=1}^{p}\cos (\pi l/q)F_{1}^{%
\text{(s)}}(2mrs_{l})  \notag \\
&&+\frac{q}{\pi }\int_{0}^{\infty }dy\frac{\sinh (y)\sinh (2qy)}{\cosh
(2qy)-\cos (q\pi )}F_{1}^{\text{(s)}}(2mr\cosh y)\bigg],  \label{T11Rreg2}
\end{eqnarray}%
where $2p\leqslant q<2p+2$. Note that the second term in the right-hand side
of this formula does not contribute to the azimuthal stress. Comparing with (%
\ref{T11Irreg}), we see that in the total regularized VEV this part is
cancelled by the part coming from the irregular mode. As a result, for the
total regularized VEV one finds the expression%
\begin{eqnarray}
\langle T_{1}^{1}\rangle _{0,\text{reg}} &=&\frac{e^{m^{2}s}}{(2\pi )^{3/2}}%
\int_{0}^{1/2s}dy\frac{y^{1/2}e^{-m^{2}/2y}}{\sqrt{1-2ys}}+\frac{m^{3}}{\pi }%
e^{m^{2}s}\bigg[\sum_{l=1}^{p}\cos (\pi l/q)F_{1}^{\text{(s)}}(2mrs_{l})
\notag \\
&&+\frac{q}{\pi }\int_{0}^{\infty }dy\frac{\sinh (y)\sinh (2qy)}{\cosh
(2qy)-\cos (q\pi )}F_{1}^{\text{(s)}}(2mr\cosh y)\bigg]+o(s).
\label{T11Reg3}
\end{eqnarray}%
The first term in the right-hand side of this expression corresponds to the
contribution coming from the Minkowski spacetime part. It is subtracted in
the renormalization procedure and for the renormalized VEV\ of the radial
stress in a boundary-free conical space one finds%
\begin{eqnarray}
\langle T_{1}^{1}\rangle _{0,\text{ren}} &=&\frac{m^{3}}{\pi }\bigg[%
\sum_{l=1}^{p}\cos (\pi l/q)F_{1}^{\text{(s)}}(2mrs_{l})  \notag \\
&&+\frac{q}{\pi }\int_{0}^{\infty }dy\frac{\sinh (y)\sinh (2qy)}{\cosh
(2qy)-\cos (q\pi )}F_{1}^{\text{(s)}}(2mr\cosh y)\bigg].  \label{T11renSp}
\end{eqnarray}%
Combining the results (\ref{T11Irreg}), (\ref{T00Rreg}), and (\ref{T11Reg3}%
), for the renormalized VEVs of the energy density and the azimuthal stress
we obtain the expressions%
\begin{eqnarray}
\langle T_{0}^{0}\rangle _{0,\text{ren}} &=&\langle T_{1}^{1}\rangle _{0,%
\text{ren}}-\frac{qm^{2}}{2\pi ^{2}r}\left[ K_{1}(2mr)-K_{0}(2mr)\right] ,
\notag \\
\langle T_{2}^{2}\rangle _{0,\text{ren}} &=&\frac{m^{3}}{\pi }\bigg\{%
\sum_{l=1}^{p}\cos (\pi l/q)F_{2}^{\text{(s)}}(2mrs_{l})  \notag \\
&&+\frac{q}{\pi }\int_{0}^{\infty }dy\frac{\sinh (y)\sinh (2qy)}{\cosh
(2qy)-\cos (q\pi )}F_{2}^{(s)}(2mr\cosh (y))\bigg\},  \label{T22renSp}
\end{eqnarray}%
where the function $F_{2}^{\text{(s)}}(x)$ is defined by Eq. (\ref{Fsi}). As
we see, for half-integer values of $\alpha $, when an irregular mode is
present, the energy density and the radial stress differ. In the case of a
massless field the radial stress is expressed as

\begin{equation}
\langle T_{1}^{1}\rangle _{0,\text{ren}}=\frac{1}{8\pi r^{3}}\bigg[%
\sum_{l=1}^{p}\frac{\cos (\pi l/q)}{s_{l}^{3}}+\frac{q}{\pi }%
\int_{0}^{\infty }dy\frac{\sinh (y)\sinh (2qy)}{\cosh (2qy)-\cos (q\pi )}%
\frac{1}{\cosh ^{3}y}\bigg],  \label{T11Sprenm0}
\end{equation}%
and for the energy density and the azimuthal stress one has:
\begin{equation}
\langle T_{0}^{0}\rangle _{0,\text{ren}}=\langle T_{1}^{1}\rangle _{0,\text{%
ren}}=-\langle T_{2}^{2}\rangle _{0,\text{ren}}/2.  \label{T00Sprenm0}
\end{equation}%
Of course, in this case the renormalized VEV is traceless.

\subsection{Region inside a circular boundary}

Now we consider the region inside a circle with radius $a$. The contribution
of the modes with $j\neq -\alpha $ is given by Eq. (\ref{TiiInt}) where now
the summation goes over $j\neq -\alpha $. For the evaluation of the
contribution coming from the mode with $j=-\alpha $, we note that the
negative-energy eigenspinor for this mode has the form \cite{Beze10b}%
\begin{equation}
\psi _{\gamma ,-\alpha }^{(-)}(x)=\frac{b_{0}}{\sqrt{r}}e^{iq\alpha \phi
+iEt}\left(
\begin{array}{c}
\frac{\gamma e^{-iq\phi /2}}{E+m}\sin (\gamma r-\gamma _{0}) \\
e^{iq\phi /2}\cos (\gamma r-\gamma _{0})%
\end{array}%
\right) ,  \label{psigamSp}
\end{equation}%
where $\gamma _{0}$ is defined by Eq. (\ref{gam0}). From boundary condition (%
\ref{BCMIT}) it follows that the eigenvalues of $\gamma $ are solutions of
the equation%
\begin{equation}
m\sin (\gamma a)+\gamma \cos (\gamma a)=0.  \label{modeqSp}
\end{equation}%
We denote the positive roots of this equation by $\gamma _{l}=\gamma a$, $%
l=1,2,\ldots $. From the normalization condition, for the coefficient in Eq.
(\ref{psigamSp}) one has%
\begin{equation}
b_{0}^{2}=\frac{E+m}{aE\phi _{0}}\left[ 1-\sin (2\gamma a)/(2\gamma a)\right]
^{-1}.  \label{b02}
\end{equation}

Using Eq. (\ref{psigamSp}), for the contributions of the mode under
consideration to the energy density and the radial stress we find:%
\begin{eqnarray}
\left\langle T_{0}^{0}\right\rangle _{j=-\alpha } &=&-\frac{q}{2\pi a^{2}r}%
\sum_{l=1}^{\infty }\frac{\gamma _{l}^{2}+\mu ^{2}+\mu \left[ \gamma
_{l}\sin \left( 2\gamma _{l}r/a\right) -\mu \cos \left( 2\gamma
_{l}r/a\right) \right] }{\sqrt{\gamma _{l}^{2}+\mu ^{2}}\left[ 1-\sin
(2\gamma _{l})/(2\gamma _{l})\right] },  \notag \\
\left\langle T_{1}^{1}\right\rangle _{j=-\alpha } &=&\frac{q}{2\pi a^{2}r}%
\sum_{l=1}^{\infty }\frac{\gamma _{l}^{2}/\sqrt{\gamma _{l}^{2}+\mu ^{2}}}{%
1-\sin (2\gamma _{l})/(2\gamma _{l})},  \label{T11Spb}
\end{eqnarray}%
where $\mu =ma$ and the presence of a cutoff function is assumed. The
contribution to the azimuthal stress vanishes: $\left\langle
T_{2}^{2}\right\rangle _{j=-\alpha }=0$. For the summation of the series in
Eq. (\ref{T11Spb}), we use the Abel-Plana-type formula \cite%
{Saha08Book,Rome02}
\begin{equation}
\sum_{l=1}^{\infty }\frac{\pi f(\gamma _{l})}{1-\sin (2\gamma _{l})/(2\gamma
_{l})}=-\frac{\pi f(0)/2}{1/\mu +1}+\int_{0}^{\infty
}dz\,f(z)-i\int_{0}^{\infty }dz\frac{f(iz)-f(-iz)}{\frac{z+\mu }{z-\mu }%
e^{2z}+1}.  \label{SumFormAp}
\end{equation}%
For the functions $f(z)$ corresponding to Eq. (\ref{T11Spb}) one has $f(0)=0$%
. The second term on the right-hand side of Eq. (\ref{SumFormAp}) gives the
part corresponding to the boundary-free geometry. As a result, the VEVs are
presented in the decomposed form (no summation)
\begin{equation}
\langle T_{i}^{i}\rangle _{j=-\alpha }=\langle T_{i}^{i}\rangle
_{0}^{(j=-\alpha )}+\langle T_{i}^{i}\rangle _{\text{b},j=-\alpha },
\label{EMTSpIn}
\end{equation}%
where the boundary-induced parts are given by the expressions%
\begin{eqnarray}
\langle T_{0}^{0}\rangle _{\text{b},j=-\alpha } &=&-\frac{q}{\pi ^{2}r}%
\int_{m}^{\infty }dx\frac{x^{2}-m^{2}+m\left[ x\sinh \left( 2xr\right)
+m\cosh \left( 2xr\right) \right] }{\sqrt{x^{2}-m^{2}}\left( \frac{x+m}{x-m}%
e^{2ax}+1\right) },  \notag \\
\langle T_{1}^{1}\rangle _{\text{b},j=-\alpha } &=&\frac{q}{\pi ^{2}r}%
\int_{m}^{\infty }dx\frac{x^{2}/\sqrt{x^{2}-m^{2}}}{\frac{x+m}{x-m}e^{2ax}+1}%
.  \label{T11Spb2}
\end{eqnarray}%
Note that%
\begin{equation}
\langle T_{0}^{0}\rangle _{\text{b},j=-\alpha }=-\langle T_{1}^{1}\rangle _{%
\text{b},j=-\alpha }+m\langle \bar{\psi}\psi \rangle _{\text{b},j=-\alpha },
\label{T00b}
\end{equation}%
where

\begin{equation}
\langle \bar{\psi}\psi \rangle _{\text{b},j=-\alpha }=\frac{q}{\pi ^{2}r}%
\int_{m}^{\infty }dx\frac{m-x\sinh (2xr)-m\cosh (2xr)}{\sqrt{x^{2}-m^{2}}%
\left( \frac{x+m}{x-m}e^{2ax}+1\right) },  \label{FCbSp}
\end{equation}%
is the corresponding part in the fermionic condensate. The contribution of
the modes $j\neq -\alpha $ remains the same and is obtained from the
corresponding expressions given above for non-half-integer values of $\alpha
$ by the direct substitution $\alpha =1/2$.

Expression (\ref{T11Spb2}) for the boundary induced part is simplified for a
massless field:%
\begin{equation}
\langle T_{0}^{0}\rangle _{\text{b},j=-\alpha }=-\langle T_{1}^{1}\rangle _{%
\text{b},j=-\alpha }=\frac{q}{48a^{2}r}.  \label{TiiSpm0}
\end{equation}%
Unlike to the fermionic condensate, the boundary induced VEVs diverge at the
circle center. Note that for a massless field these VEVs are finite on the
boundary. Adding the part corresponding to the regular modes, for the total
VEVs in the massless case we get%
\begin{eqnarray}
\langle T_{0}^{0}\rangle _{\text{b}} &=&\frac{q}{48a^{2}r}-\frac{q}{\pi
^{2}a^{3}}\sum_{n=1}^{\infty }\int_{0}^{\infty
}dx%
\,x^{2}V_{qn-1/2,qn+1/2}^{(I)}(x)[I_{qn-1/2}^{2}(xr/a)-I_{qn+1/2}^{2}(xr/a)],
\notag \\
\langle T_{2}^{2}\rangle _{\text{b}} &=&\frac{2q^{2}}{\pi ^{2}a^{2}r}%
\sum_{n=1}^{\infty }n\int_{0}^{\infty
}dx\,xV_{qn-1/2,qn+1/2}^{(I)}(x)I_{qn-1/2}(xr/a)I_{qn+1/2}(xr/a),
\label{T22Spm0}
\end{eqnarray}%
and for the radial stress one has $\langle T_{1}^{1}\rangle _{\text{b}%
}=-\langle T_{0}^{0}\rangle _{\text{b}}-\langle T_{2}^{2}\rangle _{\text{b}}$%
. The boundary-free parts in this case are given by Eqs. (\ref{T11Sprenm0})
and (\ref{T00Sprenm0}).

In the region outside a circular boundary there are no irregular modes and
the VEV of the energy-momentum tensor is a continuous function of the
parameter $\alpha $ at half-integer values. The corresponding expression is
obtained taking the limit $\alpha _{0}\rightarrow 1/2$ in the expressions
for the VEVs given above for $\alpha _{0}\neq 1/2$.

\section{Conclusion}

\label{sec:Conc}

We have investigated the VEV of the energy-momentum tensor for a massive
fermionic field in a (2+1)-dimensional conical spacetime with a circular
boundary on which the field obeys MIT bag boundary condition. In addition,
we have assumed the presence of magnetic flux located at the cone apex. A
special case of boundary conditions at the apex is considered when the MIT
bag boundary condition is imposed at a finite radius, which is then taken to
zero. In the presence of a circular boundary, the VEV of the energy-momentum
tensor is decomposed into the boundary-free and boundary-induced parts.

First we consider the geometry of a conical space without boundaries. The
corresponding VEV is evaluated by making use of mode sum formula (\ref%
{modesum}) with the eigenspinors given by Eq. (\ref{psi0}). For the
regularization of the mode sums we have introduced an exponential cutoff
function. The application of the formula (\ref{Iser2}) allowed us to
explicitly extract from the VEVs the parts corresponding to the Minkowski
spacetime in the absence of the magnetic flux. The renormalization is
reduced to the subtraction of this part. The renormalizaed VEVs in the
boundary-free geometry are given by Eq. (\ref{T11ren}). These VEVs are even
and periodic functions of the parameter $\alpha $, related to the magnetic
flux by Eq. (\ref{alfatilde}). The corresponding radial stress is equal to
the energy density. For a massless field the renormalized VEV of the energy
density is expressed as Eq. (\ref{Tiirenm0}), and the azimuthal stress is
obtained from the zero-trace condition. In the special case of integer $q$
and for the parameter $\alpha $ given by Eq. (\ref{alphaSpecial}), the
general formula is reduced to Eq. (\ref{Tii0ren}). In this case, the
renormalized VEV vanishes in a conical space with $q=2$. Various other
special cases are considered. In particular, for the magnetic flux in
background of Minkowski spacetime one has the expressions (\ref{Tii0q1}). In
this case the corresponding energy density is positive.

The effects induced by a circular boundary, concentric with the cone apex,
are considered in Sect. \ref{sec:EMTinside}. In the interior region the
eigenvalues for $\gamma $ are quantized by the boundary condition and they
are solutions of Eq. (\ref{gamVal}). The mode sums for the separate
components of the energy-momentum tensor are given by Eq. (\ref{T11Gen}) and
contain the summation over these eigenvalues. The application of the
Abel-Plana-type summation formula allows us to extract from the VEVs the
parts corresponding to the boundary-free geometry and to present the
boundary-induced parts in terms of rapidly convergent integrals suitable for
numerical evaluation. The corresponding expressions are given by Eq. (\ref%
{TiiInt}). The boundary-induced parts are even and periodic functions of the
parameter $\alpha $ with the period equal to 1. Note that for the
boundary-induced part the energy density is not equal to the radial stress.
For a massless field the general formulas are reduced to Eq. (\ref{T22bm0})
for the energy density and the azimuthal stress. The expression for the
radial stress is obtained by using the tracelessness of the energy-momentum
tensor. At the cone apex the boundary induced VEVs vanish for $|\alpha
_{0}|<(1-1/q)/2$ and they diverge when $|\alpha _{0}|>(1-1/q)/2$. In
particular, the VEVs are divergent for a magnetic flux in background of
Minkowski spacetime. Near the boundary, the boundary-induced parts in the
VEVs dominate. The leading terms in the asymptotic expansions over the
distance from the boundary are given by Eq. (\ref{Tikbound}). The energy
density is negative near the boundary, whereas the stresses may change the
sign with dependence of the field mass.

The vacuum energy-momentum tensor in the region outside a circular boundary
is considered in Sect. \ref{sec:EMToutside}. After the subtraction of the
boundary-free parts and by making use of complex rotation, we present the
boundary-induced parts in the form (\ref{T11bext}) and (\ref{T22bm0ext}),
for massive and massless fields respectively. As in the interior region,
they are even and periodic functions of the parameter $\alpha $ with the
period equal to 1. For small values of the circle radius when the radial
distance is fixed, the boundary-induced VEVs for a massive field behave as $%
(a/r)^{2q_{\alpha }}$ with $q_{\alpha }$ defined by Eq. (\ref{qalf}). For a
massless field the leading terms vanish and the behavior of the VEVs is like
$(a/r)^{2q_{\alpha }+1}$. At large distances from the circle, the boundary
induced VEVs are exponentially suppressed for a massive field and they decay
as power-law for a massless field. For points near the boundary the VEVs
diverge. The leading terms in the expansions over the distance from the
boundary are given by Eq. (\ref{T00Extnb}).

In the case of half-integer values of the parameter $\alpha $, a special
consideration is needed for the mode with $j=-\alpha $. In the boundary-free
geometry the corresponding mode functions are given by Eq. (\ref{psibetSp}).
The contribution of the special mode to the azimuthal stress vanishes. The
renormalized expressions are given by Eq. (\ref{T11renSp}) for the radial
stress and by Eq. (\ref{T22renSp}) for the energy density and azimuthal
stress. Note that, in the case under consideration the radial stress, in
general, is not equal to the energy density. This equality takes place for a
massless field only (see Eqs. (\ref{T11Sprenm0}) and (\ref{T00Sprenm0})). In
the presence of circular boundary and for half-integer values of $\alpha $,
the mode function for the special mode $j=-\alpha $ is given by Eq. (\ref%
{psigamSp}). The corresponding eigenvalues for $\gamma $ are roots of Eq. (%
\ref{modeqSp}). Similar to the boundary-free case, the contribution of the
special mode to the VEV of the azimuthal stress vanishes. The expressions
for the energy density and radial stress, Eq. (\ref{T11Spb}), are given in
terms of series over the eigenvalues of $\gamma $. For the summation of
these series we use the summation formula (\ref{SumFormAp}). This allows us
to separate the boundary-free part. The boundary-induced parts for the
contributions of the special mode are given by Eqs. (\ref{T11Spb2}) and (\ref%
{TiiSpm0}) for massive and massless fields, respectively. The total VEVs are
obtained adding the parts coming from the modes with $j\neq -\alpha $. The
latter are obtained from the formulas given before, putting directly
half-integer values for $\alpha $. In particular, for a massless field we
have the expressions (\ref{T22Spm0}).

From the point of view of the physics in the region outside the conical
defect core, the geometry considered in the present paper can be viewed as a
simplified model for the non-trivial core. This model presents a framework
in which the influence of the finite core effects on physical processes in
the vicinity of the conical defect can be investigated. The corresponding
results may shed light upon features of finite core effects in more
realistic models, including those used for defects in crystals and
superfluid helium. In addition, the problem considered here is of interest
as an example with combined topological and boundary-induced quantum
effects, in which the physical characteristics can be found in closed
analytic form.

Nanocones of carbon appear as a natural environment for applications of the
calculations presented in this paper. Like graphene, carbon nanocones, have
long-wavelength free electrons which are described effectively as Dirac
fermions. Localized defects like the apex and the boundary of the cone act
as scatterers producing standing-wave patterns in the electron density. The
interaction between these defects is given by $\langle T^{00}\rangle $ as
computed in section \ref{sec:EMTinside}. From this, the force between the
scatterers can be estimated. More importantly, our work sets the background
for the study of adsorbed atoms in the nanocone surface, a subject of very
high interest nowadays \cite{Adis11} since they are good candidates for gas
storage. Adsorbed atoms become additional defects acting, again, as electron
scatterers. Therefore, the fermionic Casimir effect with the inclusion of
extra point defects gives the interaction between the adsorbed atoms and the
apex or the boundary and among themselves \cite{Shyt09}. This will be the
subject of a forthcoming publication.

\section*{Acknowledgment}

E.R.B.M. and F.M. thank Conselho Nacional de Desenvolvimento Cient\'{\i}fico
e Tecnol\'{o}gico (CNPq) for partial financial support. A.A.S. was supported
by the Program PVE/CAPES.

\end{document}